\documentclass[useAMS,usenatbib]{mn2e}

\usepackage{color,colortbl}
\usepackage{graphicx}
\usepackage{mathptmx}

\usepackage{natbib}

\usepackage{times}
\usepackage[english]{babel}
\usepackage{footnote}
\definecolor{LightGray}{gray}{0.9}
\definecolor{LightGray1}{gray}{0.8}

\def\degr{\hbox{$^\circ$}}

\def\farcs{\hbox{$.\!\!^{\prime\prime}$}}

\title[Near-parabolic comets observed in 2006--2010]{Near-parabolic comets observed in 2006--2010.
The individualized approach to 1/a-determination and the new
distribution of original and future orbits}
\author[ M. Kr\'olikowska and P.A. Dybczy\'{n}ski ]{Ma\l gorzata Kr\'olikowska$^1$\thanks{E-mail:
mkr@cbk.waw.pl} and Piotr A. Dybczy\'{n}ski$^2$\thanks{E-mail: dybol@amu.edu.pl} \\
$^1$Space Research Centre of the Polish Academy of Sciences,
Bartycka 18A, 00-716 Warsaw, Poland \\
$^2$Astronomical Observatory Institute, Faculty of Physics,
A.~Mickiewicz University,
S\l oneczna 36, 60-286 Pozna\'{n}, Poland\\
}

\makeatother

\usepackage{babel}
\begin{document}
\pagerange{\pageref{firstpage}--\pageref{lastpage}} \pubyear{2013}

\maketitle

\label{firstpage}

\begin{abstract}
Dynamics of a complete sample of small perihelion distance
near-parabolic comets discovered in the years 2006 -- 2010 are
studied (i.e. of 22 comets of $q_{\rm osc}<3.1$\,au).

First, osculating orbits are obtained after a very careful positional
data inspection and processing, including where appropriate, the method
of data partitioning for determination of pre- and post-perihelion
orbit for tracking then its dynamical evolution. The nongravitational
acceleration in the motion is detected for 50~per cent of investigated
comets, in a few cases for the first time. Different sets of nongravitational
parameters are determined from pre- and post-perihelion data for some
of them. The influence of the positional data structure on the possibility
of the detection of nongravitational effects and the overall precision
of orbit determination is widely discussed.

Secondly, both original and future orbits were derived by means of
numerical integration of swarms of virtual comets obtained using a
Monte Carlo cloning method. This method allows to follow the
uncertainties of orbital elements at each step of dynamical
evolution. The complete statistics of original and future orbits
that includes significantly different uncertainties of 1/a-values is
presented, also in the light of our results obtained earlier. Basing
on 108~comets examined by us so far, we conclude that only one of
them, C/2007~W1~Boattini, seems to be a serious candidate for an
interstellar comet. We also found that 53~per cent of
108~near-parabolic comets escaping in the future from the Solar
system, and the number of comets leaving the Solar system as so
called Oort spike comets (i.e. comets suffering very small planetary
perturbations) is 14 per cent.

A new method for cometary orbit quality assessment is also proposed
by means of modifying the original method, introduced by \citet{mar-sek-eve:1978}.
This new method leads to a better diversification of orbit quality
classes for contemporary comets.\end{abstract}

\section{Introduction}

\label{sec:Introduction}

The origin of comets is a problem discussed for centuries but still
not fully understood. An important element of this puzzle is a
question of the observed source of near-parabolic comets. There are
two important observational facts that should help us to find an
answer. First is the almost perfectly spherically symmetric
distribution of their perihelia directions, what has lead Hal
Levison \citeyearpar{levison:1996} to call these comets
{\textit{nearly isotropic comets}} (NICs). The second is the
striking distribution of their original (i.e. before entering the
planetary zone) orbital energies, typically expressed in terms of
the reciprocal of the original semimajor axis, see for example
\citealp{fernandez_book:2005}, page~105. This distribution, highly
concentrated near zero, was first pointed out by Oort
\citeyearpar{oort:1950} and used as an evidence, that the Solar
system is surrounded by a huge, spherical cloud of comets, now
called the Oort Cloud. Oort analysed the sample of only 19 precise
original cometary orbits but since that time hundreds of such orbits
were obtained and their $1/a_{\rm {ori}}$ strong concentration in
the interval between zero and $100\times10^{-6}$au$^{-1}$ is still
evident. Since that time more and more authors call comets with the
original semimajor axis in this range the \emph{Oort spike comets}.
This term however, is a source of a serious misunderstanding since a
very popular and widely repeated opinion that ``\textit{Comets in
the spike come from the Oort cloud}'' (see for example
\citealp{fernandez_book:2005}, page~104) seems to be far-reaching
simplification of reality. As early as in 1978
\citeauthor{mar-sek-eve:1978} (hereafter MSE) formulated an opinion,
that comets from the Oort spike ``are \textit{probably} making their
first passage through the inner part of the solar system''. They
stressed (through the use of 'probably' in italic) that this is only
a guess or assumption. This word was omitted in the majority of
following papers and now is usually and incorrectly postulated that
this is an obvious fact that comets having original barycentric
semimajor axis greater than about 10\,000~au (or even a few thousand
au) are dynamically new. The simplest evidence for this to be
erratic is the fact, that a significant percentage of future (when
leaving planetary zone) semimajor axes of near-parabolic comets are
still in the spike but potential observers during next perihelion
passages cannot treat them as making their first visit among
planets. The term \textit{Solar system transparency} i.e. the
probability that a near-parabolic comet would pass through the
observability zone experiencing infinitesimal planetary
perturbations was first proposed and discussed by
\citet{dyb-trans:2004,dyb-belgrad:2005} and recently also studied by
\citealp{fouchard-r-f-v:2013}. They showed the dependence of this
probability on a perihelion distance and estimated it to vary from
almost zero for smallest perihelion distances, through 20~per cent
at $q=5$\,au up to 70\,per cent at $q=10$~au. The study of motion of
the observed large perihelion distance LPCs through the zone of
strong planetary perturbations, often called the Jupiter--Saturn
barrier, was also recently carried by present authors
\citep[hereafter Paper~2]{dyb-kroli:2011}. In the observed sample of
LPCs examined by us so far (i.e. 108~LPCs of $1/a_{{\rm
ori}}<10^{-4}$\,au$^{-1}$) we estimated the Solar system
transparency to be on the level of 14~per cent.

In the current paper we will use both terms: 'near-parabolic comets'
and 'long period comets', as well as the abbreviation LPCs, treating
them as equivalent.

There is another strong evidence that not all \textit{Oort spike
comets } make their first visit into the planetary zone. The
significant per cent of the previous perihelia obtained  from the
detailed studies of their past dynamical evolution are placed well
inside the planetary zone (e.g. Paper~2, and \citealp[hereafter
Paper~1]{kroli-dyb:2010}). Formulating his hypothesis,
\citet{oort:1950} assumed that near-parabolic comets moving on
Keplerian orbits outside the planetary perturbation zone are
sometimes (mostly near an aphelion) perturbed by passing stars.
Since that time, our knowledge on their dynamics has significantly
increased, mainly by recognizing the importance of the Galactic
perturbations in their motion (see \citealp{dones-w-l-d:2004} for a
review). Using the first order approximation, one can see that the
strength of the Galactic perturbation on a perihelion distance
scales with $a^{7/2}$ \citep{dones-w-l-d:2004}. Basing on 53
observed LPCs with $q\ge3.0$\,au we estimated this relation to be
$\Delta q\sim a^{4.06\pm0.16}$, see Paper~2 for details. While early
estimations of the Galactic disc matter density $\rho$ lead to the
conclusion that for a comet to 'jump over' the Jupiter-Saturn
barrier in one orbital period it is sufficient to have the semimajor
axis $a>10\,000$\,au, the contemporary value of $\rho=0.100$
M$_{\odot}$\,pc$^{-3}$ makes this limiting semimajor axis value much
larger, typically 20\,000 -- 28\,000\,au
\citep{levison-d-d:2001,dones-w-l-d:2004,morby:2005}.

Nowadays it is clear that information about the $1/a_{{\rm ori}}$-value
is not sufficient to determine the dynamical status of so-called Oort
spike comets and the previous perihelion distance must be inspected
as we postulate starting from the paper by \citet{dyb-hist:2001}
and what previously was discussed by \citet{yabushita:1989}. In his
paper, Yabushita showed that only 18 of 48 Oort spike comets discovered
up to 1989 are dynamically new (less than 40 per cent). However, to
know correctly the dynamical status of 'Oort spike' comet we should
follow the cometary orbit to the previous perihelion taking into account
not only the Galactic disc tide as Yabushita did in his approximation
but also account for the Galactic centre term \citep{fouchard-f-m-v:2005}
and for perturbations from passing stars \citep{fouchard-r-f-v:2011}.
Even then we might only know whether the previous perihelion passage
of an actual comet was inside the inner part of the Solar system or
beyond the planetary zone.

Starting from the classical paper of MSE it becomes clear that an
another important factor, namely nongravitational (NG) effects,
should be taken into account in the context of the determination of
original inverse semimajor axes of near-parabolic comets. As it was
already demonstrated by
\citet{krolikowska:2001,krolikowska:2002,krolikowska:2004} the
NG~accelerations should be included when determining osculating
orbits of LPCs, since they can significantly change their original
semimajor axes. This effect was also clearly presented in Paper~1.

Therefore, the aim of this paper is twofold.

First, we develop our methods of a precise osculating orbit
determination (hereafter a nominal orbit) for the purpose of
previous and next perihelion passage calculations, that will be
described in the second part of this investigation
(\citealp[hereafter Part~II]{dyb-krol:2013}). Here, we try to
determine an NG~orbit for each investigated here LPC (Section
\ref{sec:Observations-and-orbit}) and next we discuss these results
with the \textit{a priori} possibilities of NG~determinations based
on the data structure (Section~\ref{sec_cometary_cases}). For a
complete sample of near-parabolic comets observed in 2006~--~2010,
and having $q<3.1$\,au and $1/a_{{\rm
ori}}<150\times10^{-6}$\,au$^{-1}$, we notice the 50~per cent of
success for the detection of NG~effects in comet's motion using the
positional data. Additional result of our study is a new method of
cometary orbit quality assessment that is described in details in
Section~\ref{sec_orbit_accuracy}.

Secondly, to know the dynamical status in the context of the previous
perihelion passage of analysed here comets, we construct a swarm of
osculating virtual comets (hereafter VCs) on the basis of the
nominal orbit derived previously for each comet. In this part of
investigation, we follow each swarm backward and forward to a
distance of 250\,au from the Sun. Thus, we obtain the original and
future orbits for each comet together with the uncertainties of all
orbital parameters. The method is described in
Section~\ref{sec_orbit_original_future}.  In that section, we focus
on two different issues: (i) the change of original inverse
semimajor axis due to incorporating the NG~acceleration in the
osculating orbit determination from the data, and (ii) the
statistics of original and future $1/a$-distribution of the sample
of 108~near-parabolic comets studied by us here and in our previous
papers. In the second aspect, we concentrate on a detailed
discussion of original and future $1/a$-distribution as well as the
observed planetary perturbation distribution in the context of all
so-called Oort spike comets studied by us so far (in Paper~1 \& 2).
Here, we use the term \textit{Oort spike} because of its popularity.
However, we always have in mind \textit{near-parabolic comets},
remembering that they consist of two populations of dynamically new
and dynamically old comets. We will return to this aspect in
Part~II.

In Part~II, we will follow each swarm of barycentric orbits of VCs
(taken at 250\,au from the Sun) to the previous and next perihelion
taking into account the Galactic perturbations and perturbations of
all known stars. Then, we will discuss the observed distribution of
Oort spike together with the problem of cometary origin. This paper
is in preparation.

\begin{savenotes}
\begin{table*}
\begin{centering}
\caption{\label{tab:Obs-mat}Description of observational material of 22~LPCs
discovered during the period 2006--2010 (columns $[4]-[8]$) and global
characteristics of orbits determined here from the entire data intervals
(columns $[9]-[10]$). Second and third columns show an osculating
perihelion distance and perihelion time, respectively. Data distribution
relative to a perihelion passage is presented in columns $[7]$ \&
$[8]$, where 'pre!' ('post!') means that all observations were taken
before (after) perihelion passage; 'pre+' ('post+') means that we
noticed significantly more pre-perihelion (or post-perihelion) measurements,
and additional '$+$' indicates the drastic dominance of pre-perihelion
(or post-perihelion) measurements in both the number and the time
interval. Column $[10]$ shows the type of the best model possible
to determine from the full interval of data (subscript 'un' informs
that for a given comet the GR and even the NG~model determined from
the entire interval of data is not satisfactory), and column $[9]$
gives the resulting orbital class determined according to Q$^{*}$
given in column {[}12{]} and recipe given in Section~\ref{sec_orbit_accuracy};
notice that the subscript 'un' in {[}10{]} means that due to an inappropriate
model the orbital class in a given case is also very preliminary.
Column $[11]$ shows our division of investigated comets into four groups
(a detailed description is in Section~\ref{sec_cometary_cases}.) }

\par\end{centering}

\centering{}{\setlength{\tabcolsep}{2.0pt} %
\begin{tabular}{lcccrccccccc}
\hline
Comet  & q$_{osc}$  & T  & Observational arc  & No  & Data  & Heliocentric  & Data  & Orbital  & Type of  & Comet  & Q{$^{*}$} \tabularnewline
name  &  &  & dates  & of  & arc span  & distance span  & type  & class  & model  & group  & eq~\ref{eq:orbit_accuracy_3} \tabularnewline
 & {[}au{]}  & {[}yyyymmdd{]}  & {[}yyyymmdd -- yyyymmdd{]}  & obs  & {[}yr{]}  & {[}au{]}  &  &  &  &  & \tabularnewline
$[1]$  & $[2]$  & $[3]$  & $[4]$  & $[5]$  & $[6]$  & $[7]$  & $[8]$  & $[9]$  & $[10]$  & $[11]$  & $[12]$\tabularnewline
\hline
C/2006 HW$_{51}$ ~Siding Spring  & 2.266  & 20060929  & 20060423 -- 20070807  & 187  & 1.3  & 2.87 -- 4.04  & full  & 1a  & GR  & C  & 7.5 \tabularnewline
\rowcolor{LightGray} C/2006 K3 ~McNaught  & 2.501  & 20070313  & 20060522 -- 20080126  & 207  & 1.7  & 3.95 -- 4.13  & ~~pre+  & 1a  & NG  & B  & 7.5 \tabularnewline
C/2006 L2 ~McNaught  & 1.994  & 20061120  & 20060614 -- 20070707  & 408  & 1.1  & 2.74 -- 3.31  & full  & 1a  & GR  & C  & 7.5 \tabularnewline
\rowcolor{LightGray} C/2006 OF$_{2}$ ~Broughton  & 2.431  & 20080915  & 20060623 -- 20100511  & 4917  & 3.9  & 7.88 -- 6.31  & full  & ~~1a+  & NG  & A  & 8.5 \tabularnewline
\rowcolor{LightGray} C/2006 P1 ~McNaught  & 0.171  & 20070112  & 20060807 -- 20070711  & 341  & 0.9  & 2.74 -- 3.34  & full  & 1b  & NG  & B  & 6.5 \tabularnewline
\rowcolor{LightGray} C/2006 Q1 ~McNaught  & 2.764  & 20080703  & 20060820 -- 20101017  & 2744  & 4.2  & 6.83 -- 7.91  & full  & ~~1a+  & NG  & A  & 9.0 \tabularnewline
\rowcolor{LightGray} C/2006 VZ$_{13}$ ~LINEAR  & 1.015  & 20070810  & 20061113 -- 20070814  & 1173  & 0.7  & 3.84 -- 1.02  & ~~~pre++  & 1b  & ~~~NG$_{{\rm un}}$  & D  & 6.5 \tabularnewline
\rowcolor{LightGray} C/2007 N3 ~Lulin  & 1.212  & 20090110  & 20070711 -- 20110101  & 3951  & 3.2  & 6.38 -- 7.83  & full  & ~~1a+  & ~~~NG$_{{\rm un}}$  & A  & 8.5 \tabularnewline
C/2007 O1 ~LINEAR  & 2.877  & 20070603  & 20060402 -- 20071113  & 183  & 1.6  & 4.99 -- 2.91  & ~~~~~post++  & 1a  & GR  & C  & 7.5 \tabularnewline
C/2007 Q1 ~Garradd  & 3.006  & 20061211  & 20070821 -- 20070914  & 43  & 24d  & 3.88 -- 4.02  & ~~~post!  & 3a  & GR  & D  & 3.5 \tabularnewline
\rowcolor{LightGray} C/2007 Q3 ~Siding Spring  & 2.252  & 20091007  & 20070825 -- 20110925  & 1368  & 4.0  & 7.64 -- 7.24  & full  & ~~1a+  & ~~~NG$_{{\rm un}}$  & A  & 9.0 \tabularnewline
\rowcolor{LightGray} C/2007 W1 ~Boattini  & 0.850  & 20080624  & 20071120 -- 20081217  & 1703  & 1.2  & 3.33 -- 2.84  & full  & 1a  & ~~~NG$_{{\rm un}}$  & B  & 7.5 \tabularnewline
\rowcolor{LightGray} C/2007 W3 ~LINEAR  & 1.776  & 20080602  & 20071129 -- 20080908  & 212  & 0.8  & 2.89 -- 2.17  & ~~pre+  & 1b  & NG  & B  & 6.5 \tabularnewline
\rowcolor{LightGray} C/2008 A1 ~McNaught  & 1.073  & 20080929  & 20080110 -- 20100117  & 937  & 2.0  & 3.73 -- 5.82  & full  & 1a  & ~~~NG$_{{\rm un}}$  & B  & 8.0 \tabularnewline
C/2008 C1 ~Chen-Gao  & 1.262  & 20080416  & 20080130 -- 20080528  & 815  & 0.3  & 1.71 -- 1.41  & ~~~~pre++  & 2a  & GR  & D  & 6.0 \tabularnewline
C/2008 J6 ~Hill  & 2.002  & 20080410  & 20080514 -- 20081207  & 390  & 0.6  & 2.04 -- 3.41  & ~~~post!  & 1b  & GR  & C  & 7.0 \tabularnewline
C/2008 T2 ~Cardinal  & 1.202  & 20090613  & 20081001 -- 20090909  & 1345  & 0.9  & 3.60 -- 1.78  & ~~pre+  & 1b  & GR  & C  & 7.0 \tabularnewline
C/2009 K5 ~McNaught  & 1.422  & 20100430  & 20090527 -- 20111028  & 2539  & 2.4  & 4.35 -- 6.25  & full  & ~~1a+  & ~GR$_{{\rm un}}$  & B  & 8.5 \tabularnewline
C/2009 O4 ~Hill  & 2.564  & 20100101  & 20090730 -- 20091214  & 785  & 0.4  & 3.04 -- 2.57  & ~pre!  & 1b  & GR  & C  & 6.5 \tabularnewline
\rowcolor{LightGray} C/2009 R1 ~McNaught  & 0.405  & 20100702  & 20090720 -- 20100629  & 792  & 0.9  & 5.06 -- 0.41  & pre!  & 1b  & NG  & B  & 7.0 \tabularnewline
C/2010 H1 ~Garradd  & 2.745  & 20100618  & 20100219 -- 20100702  & 47  & 0.2  & 2.82 -- 2.75  & full  & 2b  & GR  & D  & 5.0 \tabularnewline
C/2010 X1 ~Elenin  & 0.482  & 20110910  & ~~20101210 -- 20110731$^{1}$  & 2254  & 0.6  & 4.22 -- 1.04  & pre!  & 1b  & ~~GR$_{{\rm un}}$  & B  & 7.0 \tabularnewline
\hline
\end{tabular}} {\footnotesize $^{1}$ ~~Comet was observed to 7~September,
however comet started to disintegrating in August. Thus, the data
were taken to the end of July.}
\end{table*}

\end{savenotes}

\section{Observations and osculating orbit determination}

\label{sec:Observations-and-orbit}

We selected all near-parabolic comets discovered during the period
2006--2010 that have small perihelion distances, i.e. $q_{\rm osc}<3.1$\,au,
and $1/a_{\rm ori}<0.000150$\,au$^{-1}$. During the same period
23~comets of $q_{\rm osc}\geq3.1$\,au and $1/a_{\rm ori}<0.000150$\,au$^{-1}$
were detected; five of them were studied in Paper~2
and nine were still observable in November 2012. This means that data
sets of these large perihelion comets were incomplete at the moment
of this investigation. Therefore, in this study we restricted to complete
sample of comets with small perihelion distances.

All results presented in this paper are based on positional data retrieved
from the IAU~Minor Planet Center in August 2012, except the case
of comet C/2010~H1 where we updated the observational data in January
2013 because then four new pre-discovery observations were published
at the Web for this comet. Global characteristics of the observational
material are given in columns 2--8 of Table~\ref{tab:Obs-mat}. Most
of comets in the investigated sample were observed on both orbital
legs (compare columns 3, 4 and 8), except of five objects. Two of
these comets (C/2007~Q1 and C/2008~J6) were discovered after perihelion
passage and three (C/2009~O4, C/2009~R1 and C/2010~X1) were not
observed after perihelion passage. The last two comets have passed
their perihelia close to the Sun at a distance of 0.40\,au and 0.48\,au,
respectively. Comet C/2010~X1 started to disintegrate about one month
before perihelion whereas C/2009~R1 was lost after perihelion. We
can suspect that C/2009~R1 also did not survive perihelion passage.
One can see that for two other comets, C/2006~VZ$_{13}$ ($q_{\rm osc}=1.01$\,au)
and C/2008~C1 ($q_{\rm osc}=1.26$\,au), the observations stopped
shortly after perihelion passage at the distance of 1.02\,au and
1.41\,au from the Sun, respectively. Thus, also in these two cases,
especially for C/2006~VZ$_{13}$ where the last observation was taken
when comet was only about 1\,au from the Earth, we can make a guess
about their possible break-up.

Comets passing close to the Sun in their perihelia are of special
interest because we should suspect detectable influence of NG~forces
on their motion. It means, however, that the orbit determination for
these LPCs is significantly more complicated than for LPCs with large
perihelion distances.

The determination of the NG~parameters in the motion of LPCs (see
Section~\ref{sub_NGmotion}) is much more difficult than in the motion
of short-period comets mainly due to limited observational material
covering one apparition or even just a half apparition, as we have
for six comets mentioned above (compare also columns 3,4 and 8 of
Table~\ref{tab:Obs-mat}). We discussed earlier in Paper~1 that
the appropriate processing of astrometric data is very important for
this purpose. In particular, the data weighting is crucial for the
orbit fitting not only for comets discovered a long time ago but also
for currently observed comets. Thus, we adopted here the same, advanced
data treatment as in our previous papers. The detailed procedure of
weighting is described in Paper1. In this procedure, each individual
set of astrometric data has been processed (selected and weighted)
during the determination of a pure gravitational orbit (GR) or NG~orbit,
independently.

\begin{savenotes}
\begin{table*}
\begin{centering}
\caption{\label{tab:NG-parameters}NG~parameters derived in orbital solutions based on entire data intervals. 
First row of each object presents NG~model given also in Table~\ref{tab:models}. Remaining models are here only to show
alternative NG~solutions (ignoring normal component of NG~acceleration, column $[4]$, and/or assuming the time shift of 
$g(r)$-function relative to perihelion passage, column $[5]$). 
Only for C/2006~K3, C/2006~OF$_2$, C/2006~P1, C/2006~Q1, C/2007~W3 and C/2009~R1 NG~models given in the first rows 
are used as preferred models for studying the dynamical evolution; compare with Table~\ref{tab:models}.
}

\par\end{centering}

\centering{}{\setlength{\tabcolsep}{11.0pt} %
\begin{tabular}{@{}lc@{$\pm$}cc@{$\pm$}cc@{$\pm$}cc@{$\pm$}ccc@{$\pm$}c@{}}
\hline
{Comet}  & \multicolumn{6}{c}{NG parameters defined by Eq.~\ref{eq:ng_std} in units of 10$^{-8}\,$au\,day$^{-2}$ }  & \multicolumn{2}{c}{$\tau$}  & {rms } & \multicolumn{2}{c}{1/a$_{\rm ori}$}  \\
{ }      & \multicolumn{2}{c}{A$_1$}  & \multicolumn{2}{c}{A$_2$}  & \multicolumn{2}{c}{A$_3$}  & \multicolumn{2}{c}{$[$ days$]$}  & {$[$arcsec$]$ } & \multicolumn{2}{c}{$[$10$^{-6}$\,au$^{-1}]$}  \\
{$[1]$}  & \multicolumn{2}{c}{{$[2]$}} & \multicolumn{2}{c}{{$[3]$}}& \multicolumn{2}{c}{{$[4]$}}&\multicolumn{2}{c}{{$[5]$}}& {$[6]$} & \multicolumn{2}{c}{{$[7]$}} \\
\hline
                     C/2006 K3         & 15.69    & 1.67    &  2.25    & 2.45    &$-$0.209   & 0.576          & \multicolumn{2}{c}{--}   & 0.54 & 61.02 & 4.63 \\
                                       & 15.66    & 1.67    &  2.40    & 2.41    &  \multicolumn{2}{c}{0.0}   & \multicolumn{2}{c}{--}   & 0.55 & 61.74 & 4.25 \\
                     C/2006 OF$_{2}$   & 2.384    & 0.168   &$-$1.370  & 0.131   &$-$0.0059  & 0.0347         & \multicolumn{2}{c}{--}   & 0.36 & 21.21 & 0.49 \\
                                       & 2.389    & 0.166   &$-$1.372  & 0.131   & \multicolumn{2}{c}{0.0}    & \multicolumn{2}{c}{--}   & 0.36 & 21.21 & 0.49 \\
                     C/2006 P1         & 0.1329   & 0.0335  & 0.03138  & 0.00397 & \multicolumn{2}{c}{0.0}    & \multicolumn{2}{c}{--}   & 0.25 & 57.17 & 4.03 \\
                     C/2006 Q1         & 33.504   & 0.700   & 1.916    & 0.550   & 10.604   & 0.189           & \multicolumn{2}{c}{--}   & 0.37 & 51.08 & 0.48 \\
                                       & 31.794   & 0.752   &$-$2.710  & 0.728   & 10.199   & 0.189           & 37.6     & 4.2           & 0.37 & 49.69 & 0.47 \\
                                       & 27.519   & 0.840   &$-$11.592 & 0.632   & \multicolumn{2}{c}{0.0}    & \multicolumn{2}{c}{50}   & ~~0.46$^1$ & \multicolumn{2}{c}{44.28} \\
                     C/2006 VZ$_{13}$  & 1.874    & 0.804   &$-$0.866  & 0.483   & 0.528    & 0.404           & \multicolumn{2}{c}{--}   & ~~0.39$^2$ & 13.96 & 4.80 \\
                                       & 1.434    & 0.686   &$-$0.547  & 0.376   & \multicolumn{2}{c}{0.0}    & \multicolumn{2}{c}{--}   & ~~0.39$^2$ & 15.86 & 4.46 \\
                                       & 4.576    & 0.115   &$-$3.041  & 0.135   & 1.220    & 0.074           & \multicolumn{2}{c}{--}   & 0.51 &$-$18.39 & 3.87 \\
                                       & 3.1277   & 0.0780  &$-$1.1912 & 0.9796  & \multicolumn{2}{c}{0.0}    & \multicolumn{2}{c}{--}   & 0.54 &$-$23.09 & 4.12\\
                     C/2007 N3         & 0.09377  & 0.00962 &$-$0.00739& 0.00611 &$-$0.12700 & 0.00145        & \multicolumn{2}{c}{--}   & 0.35 & 32.77 & 0.18 \\
                                       & 0.08678  & 0.00814 &$-$0.02141& 0.00696 &$-$0.13334 & 0.00190        &  11.3    & 1.9           & 0.35 & 32.39 & 0.17 \\
                                       & 0.08650  & 0.01117 &$-$0.01535& 0.00961 & \multicolumn{2}{c}{0.0}    & \multicolumn{2}{c}{11.3} & ~~0.49$^1$ & \multicolumn{2}{c}{32.59} \\
                     C/2007 Q3         & 0.156    & 0.180   & 2.675    & 0.103   & 1.657     & 0.037          & \multicolumn{2}{c}{--}   & 0.39 & 39.13 & 0.49 \\
                                       & 0.114    & 0.239   & 2.086    & 0.136   & \multicolumn{2}{c}{0.0}    & \multicolumn{2}{c}{--}   & 0.48 & 36.61 & 0.64 \\
                                       & 0.014    & 0.189   & 2.589    & 0.089   & 1.592     & 0.037          &$-$25.2   & 5.1           & 0.39 & 40.78 & 0.56 \\
                                       &$-$0.454  & 0.230   & 2.080    & 0.118   & \multicolumn{2}{c}{0.0}    & \multicolumn{2}{c}{$-$25}& ~~0.48$^1$ & \multicolumn{2}{c}{36.90} \\
                     C/2007 W1         & 3.9442   & 0.0125  &$-$0.6133 & 0.0175  &$-$0.06023 & 0.00361        & 22.32    & 4.6           & 0.67 &$-$36.56~~ & 1.86 \\
                                       & 4.0627   & 0.0193  &$-$0.9758 & 0.0170  &$-$0.05387 & 0.00433        & \multicolumn{2}{c}{--}   & 0.96 &$-$82.30~~ & 2.61 \\
                     C/2007 W3         & 4.968    & 0.572   & 2.248    & 0.581   &$-$1.084   & 0.316          & \multicolumn{2}{c}{--}   & 0.52 & 31.38 & 3.85 \\
                                       & 4.500    & 0.614   & 1.822    & 0.643   &$-$0.655   & 0.503          &$-$22     & 21            & 0.52 & 30.70 & 6.72 \\
                                       & 5.458    & 0.473   & 0.957    & 0.279   & \multicolumn{2}{c}{0.0}    & \multicolumn{2}{c}{$-$22}& ~~0.52$^1$ & \multicolumn{2}{c}{25.71} \\
                     C/2008 A1         & 5.5964   & 0.0570  & 0.7136   & 0.0384  & 0.16800   & 0.00815        & 5.76    & 0.6            & 0.44 &123.07 & 1.50 \\
                                       & 5.1495   & 0.0320  & 0.9915   & 0.0201  & 0.19393   & 0.00763        & \multicolumn{2}{c}{--}   & 0.45 &120.14 & 1.56 \\
                                       & 5.9732   & 0.0362  & 0.3252   & 0.0216  & \multicolumn{2}{c}{0.0}    & \multicolumn{2}{c}{10}   & ~~0.49$^1$ & \multicolumn{2}{c}{113.73} \\
                     C/2009 R1         & 5.798    & 0.490   &$-$1.418  & 0.359   & 0.776     & 0.207          & \multicolumn{2}{c}{--}   & 0.51 & 12.16 & 3.29 \\
\hline
\end{tabular}} 

{\footnotesize ~~~~~~$^{1}$~~{Best fitting asymmetric NG~models (in the sense of minimal value of rms) with assumed A$_3$=0.0)},
\newline $^{2}$ ~~{NG~models based on pre-perihelion data only (see also Table~\ref{tab:models}).~~~~~~~~~~~~~~~~~~~~~~~~~~~~~~~~~~~~~~~~~~~~~~~~~~~~~~~~~~~~~~~~~~~~~~~~~~~~}}
\end{table*}

\end{savenotes}

\subsection{The non-gravitational acceleration in the comet's motion}

\label{sub_NGmotion}

To determine the NG~cometary orbit we used the standard formalism
proposed by \citet[hereafter MSY]{marsden-sek-ye:1973} where the
three orbital components of the NG~acceleration acting on a comet
are scaled with a function $g(r)$ symmetric relative to perihelion:

\begin{eqnarray}
F_{i}=A_{\rm i}\cdot & g(r) & ,\qquad A_{{\rm i}}={\rm ~const~~for}\quad{\rm i}=1,2,3,\label{eq:g_r}\\
 & g(r) & =\alpha\left(r/r_{0}\right)^{-2.15}\left[1+\left(r/r_{0}\right)^{5.093}\right]^{-4.614},\label{eq:ng_std}
\end{eqnarray}

\noindent where $F_{1},\, F_{2},\, F_{3}$ represent the radial, transverse
and normal components of the NG~acceleration, respectively and the
radial acceleration is defined outward along the Sun-comet line. 
The normalization constant $\alpha=0.1113$ gives $g(1$~AU$)=1$;
the scale distance $r_{0}=2.808$~AU. From orbital calculations,
the NG~parameters $A_{1},A_{2}$, and $A_{3}$ were derived together
with six orbital elements within a given time interval (numerical
details are described in \citealp{krolikowska:2006a}). The standard
NG~model assumes that water sublimates from the whole surface of
an isothermal cometary nucleus. The asymmetric model of NG~acceleration
is derived by using $g(r(t-\tau))$ instead of g(r(t)). Thus, this
model introduces an additional NG~parameter $\tau$ -- the time displacement
of the maximum of the $g(r)$ relative to the moment of perihelion
passage.

Typically, the radial component, $A_{1}$, derived in a symmetric
model is positive (reflecting, in average, the stronger sublimation
of this part of cometary surface that is directed to the Sun) and
dominate in magnitude over the transverse and normal components.
This model, however, does not include the possibility of location of
an active region(s) on cometary surface. The negative radial
component, $A_{1}<0$, derived in this model, would give a first
indication for the asymmetric model of g(r)~function (with rather
large time displacement of a maximum of the $g(r)$ relative to
perihelion) or for the existence of active region(s) on comet's
surface. Described model is very successful in representing the
astrometric data, thus also in allowing the realistic dynamical
evolution predictions. However, this NG~force model does not
represent an accurate representation of the actual processes taking
place in the cometary nucleus (e.g. see \citealp{yeomans:1994}).
Thus, using this standard model of NG~acceleration we can only go as
far as a very general and a very qualitative discussion in this
field. Therefore, in this paper we place a strong emphasis on
orbital dynamics of near-parabolic comets examined here, accounting
only for evident physical events, like disruption or fragmentation.
For example, in all such cases registered, we notice the coincidence
between bursts (or disruptions) and anomalies occurring in
O-C-diagram. Therefore, we decided to exclude such data intervals in
the process of osculating orbit determination.

In the present investigation we decided to use an NG~force model
with the smallest number of parameters needed. We tested asymmetric
model for several comets from the sample studied here but no improvement
was observed. Moreover, in two cases, C/2006~HW$_{51}$ and C/2006~P1,
just two NG~parameters were determinable with reasonable accuracy.

For comets with long time sequences of astrometric data (e.g. belonging
to comet group A -- see column 11 of Table~\ref{tab:Obs-mat}) we
also tested a more general form of the dependence of NG~acceleration
on a heliocentric distance:
\begin{eqnarray}
F_{i}=A_{{\rm i}}^{*}\cdot h(r),\qquad A_{{\rm i}}^{*}={\rm ~const~~for}\quad{\rm i}=1,2,3.\label{eq_nonstandard}
\end{eqnarray}
where we adopted two different form for a dependence of acceleration
on a heliocentric distance, $h(r)$, namely: more general $g(r)$-like
function, $g^{*}(r)$ (hereafter GEN model type), and Yabushita function, $f(r)$, based on the
carbon monoxide sublimation rate \citep[hereafter YAB model type]{yabushita:1996}.

Thus, we consider here also the following two types of NG~models:
\begin{itemize}
\item GEN:
\begin{equation}
h(r)=g^{*}(r)=\alpha\left(r/r_{0}\right)^{-m}\left[1+\left(r/r_{0}\right)^{n}\right]^{-k}\label{eq_NG_model:GEN}
\end{equation}

\item YAB:
\begin{equation}
h(r)=f(r)=\frac{1.0006}{r^{2}}\times10^{-0.07395(r-1)}\cdot\left(1+0.0006r^{5}\right)^{-1}\label{eq_NG_model:YAB}
\end{equation}

\end{itemize}
In a GEN type of NG~model we have additional four free parameters:
scale distance, $r_{0}$, and exponents $n,m,k$. The function $g^{*}(r)$
is normalized similarly as standard $g(r)$ function, thus $\alpha$
is calculated from the condition: $g^{*}({\rm 1au})=1$, also the
$f(r)$ function was normalized to unity at $r=1$\,au.

In contrast to comets C/2002~T2 and C/2001~Q4 investigated by \citet[hereafter Paper~3]{kroli-dyb:2012},
we found that GEN and YAB types of NG~model did not improve the orbital
data fitting for investigated here comets with more than 3\,yr interval
of data covering the wide range of heliocentric distances (C/2006~OF$_{2}$,
C/2006~Q1, C/2007~N3 and C/2007~Q3).

We were able to determine the NG~effects for 11 of 22~comets
discovered in the period 2006--2010 (see next section). As far as we
know this is the richest sample of LPCs with NG~effects of currently
(in February 2013) published in periodicals and on the Web Pages.
Many sources of osculating orbits of comets are available only in
the Web. From these sources we noticed that \citet{NakanoWeb} and
\citet{IMCCE_Web} published NG~orbits for largest per cents of
comets in comparison to other Web~sources. In February~2013 both
sources presented NG~osculating orbits for six comets from the
period of 2006--2010, whereas we determined NG~orbits for eleven
comets discovered in this period. Additionally, only at Nakano page
\citeyearpar{NakanoWeb} and at \citet{IAU_MPC_Web} values of
original and future $1/a$ are given; in the second Web source for
three comets with NG~orbits: C/2007~W1, C/2008~A1 and C/2008~T2.
More details about NG~models derived in the present studies are
given in the next two sections. In Section
\ref{sec_orbit_original_future} we discuss the change of the
$1/a_{\rm ori}$ due to incorporation of the NG~acceleration in the
process of osculating orbit determination using the positional data.

\subsection{Osculating orbit determination from the full data interval}
\label{sub_orbit_full_data}

\begin{figure}
\includegraphics[width=8.8cm]{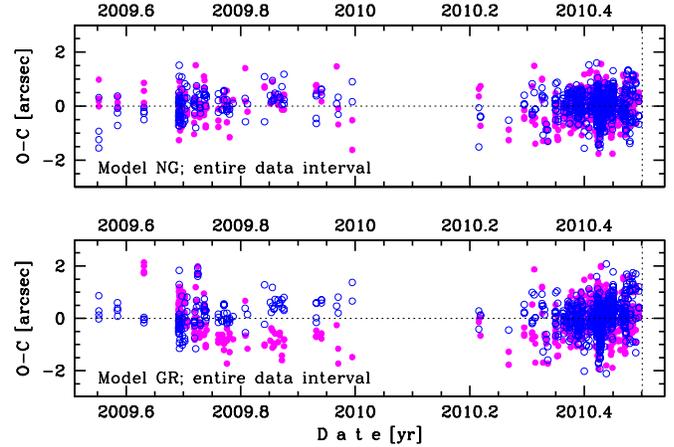}

\caption{The O-C diagrams for comet C/2009~R1~McNaught (weighted
data). The upper figure shows O-C based on an NG~solution whereas
the lower figure presents O-C based on a pure GR~orbit. Residuals in
right ascension are shown as magenta dots and in declination -- as
blue open circles; the moment of perihelion passage is shown by
dashed vertical line.}
\label{fig:OC_09r1}
\end{figure}

\begin{figure}
\includegraphics[width=8.8cm]{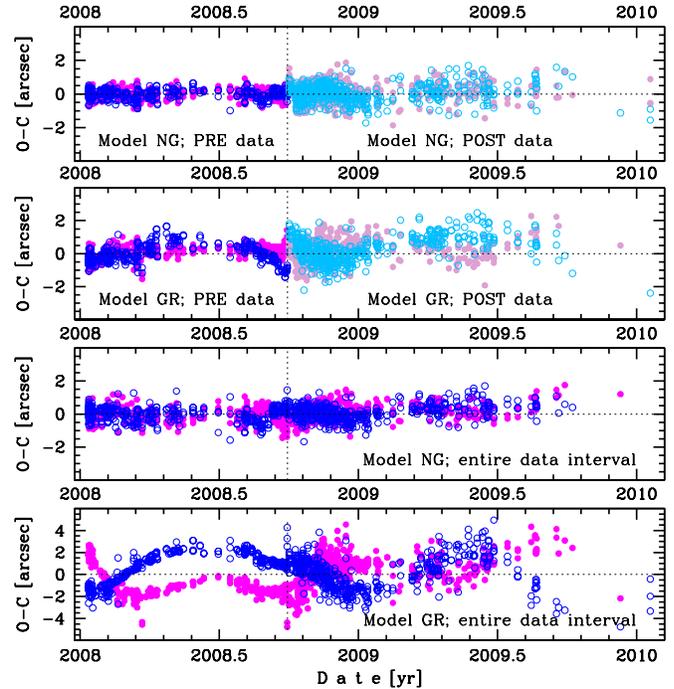}

\caption{The O-C diagrams for comet C/2008~A1~McNaught (weighted
data). Two lower panels show O-C based on the entire data interval
for an NG~orbit and a pure GR~orbit, respectively. Two upper panels
present the O-C~diagrams for orbits (NG or GR) determined
individually for pre-perihelion and post-perihelion orbital
branches. Residuals in right ascension are shown as magenta or plum
dots and in declination -- as blue or light blue open circles; the
moment of perihelion passage is shown by dashed vertical line.}
\label{fig:OC_08a1}
\end{figure}

We found that NG~accelerations are well-detectable in the motion of
eleven comets during their periods covered by positional
observations. These are comets C/2006~K3, C/2006~OF$_{2}$,
C/2006~P1, C/2006~Q1, C/2006 VZ$_{13}$, C/2007~N3, C/2007~Q3,
C/2007~W1, C/2007~W3, C/2008~A1, C/2009~R1. 

These models are shown as coloured light grey rows in Table~\ref{tab:Obs-mat}, whereas the
values of original and future 1/a for these models are given in
Table~\ref{tab:models}, except comet C/2006~VZ$_{\rm 13}$ where only
the model based on pre-perihelion data are shown for the reason
discussed in Section~\ref{sub_cometary_case_D}. 

The NG~parameters of these symmetric NG~models are given in Table~\ref{tab:NG-parameters} 
in the first row of each individual comet. Additionally, in this table we presented some alternate models
that we found as less certain in the sense of orbital fitting to data (three creteria are given just below). 
Asymmetric NG~solution (with $\tau$-parameter, see previous section) was possible to determine for six comets: C/2006 Q1, C/2007 N3, C/2007 Q3 C/2007 W1, C/2007 W3 and C/2008 A1 (see Table~\ref{tab:NG-parameters}). However, only in the case of C/2007 W1 we noticed substantial improvement of orbital fitting to data and in the case of C/2008~A1 -- the infinitesimal adjustment (in the sense of at least one of three criteria given in section 2.2). For that reason both asymmetric model are also given in Table~\ref{tab:models} as the best NG~solution derived from the entire data set. 

It was surprising that in some cases a normal component od NG~acceleration improves
the orbit's fitting to data to the much higher degree than the $\tau$, see the NG~solutions 
for C/2006~Q1, C/2007~N3 and C/2007~Q3. In another words, neglecting the 
normal component A$_3$ and determining the A$_1$, $A_2$ and $\tau$ we get the orbital solution with significantly worse data fitting. 

One can noticed that almost all these alternate models give original 1/a in a very
good agreement with original 1/a-values based on NG~models chosen by us to dynamical evolution investigation. 
The only exception is C/2007~W1 where symmetric model give more hiperbolic orbit that the asymmetric model
with $\tau$, however both original 1/a-values are certainly negative. However, this comet exhibits erratic behaviour 
and its more appropriate solutions (given in the Tables~\ref{tab:models} and \ref{tab:NGTwoComets}) are discussed later. Comet C/2006~VZ$_{13}$ is quite a different case because of the almost exclusively pre-perihelion data and some possibility of disintegrating processes close to perihelion.

By 'well-detectable' NG~effects we mean that assuming standard NG~model of
motion (see Section~\ref{sub_NGmotion}) we noticed (for each of
these eleven comets) the better orbit fitting to data in comparison
to a pure GR~orbit measured by three criteria:
\begin{itemize}
\item decrease in rms,
\item overcoming or reducing the improper trends in O-C~time variations,
\item increasing the similarity of the O-C distribution to a normal distribution.
\end{itemize}
More details and examples how this analysis works are given in Papers1--3,
therefore only two examples are given below.

Figs~\ref{fig:OC_09r1}--\ref{fig:OC_08a1} show the comparison
between the O-C~diagrams of NG~orbit and GR~orbit for C/2009~R1 with
moderately manifesting NG~effects in the motion, and C/2008~A1 with
spectacularly visible NG~effects in the motion, respectively. We
additionally noticed the decrease in rms from 0\farcs63 (GR~orbit)
to 0\farcs51 (NG~orbit) for C/2009~R1 and from 1\farcs44 (GR~orbit)
to 0\farcs44 (NG~orbit) for C/2008~A1. One can see in
Fig.~\ref{fig:OC_09r1} that trends easily visible in the O-C~diagram
based on GR~orbit disappear for NG~orbit. Moreover at the beginning
of the data set, there are four observations taken on 2009~07~20 and
four in 2009~08~01. According to our selection procedure all these
measurements in right ascension are not used for GR~orbit
determination due to unacceptable large residuals whereas in the
case of NG~orbit all are well-fitted as one can see in the upper
panel of Fig.~\ref{fig:OC_09r1}. Such a data recovery, in particular
at the edges of observational period (what we noticed in many cases
of the NG~orbit determination) is an additional argument for a
prevalence of NG~orbit (and NG~model). Comet C/2008~A1 is a very
special case because we still can see in the O-C~diagram
well-visible trends in residuals in right ascension and declination
for NG~orbit determined from a whole data set (third panel from the
top in Fig~\ref{fig:OC_08a1}). Also, the distributions of residuals
in $\alpha$ and $\delta$ are not fitted well to normal
distributions. Thus, for this comet it was necessary to divide the
data into two parts: data before and after perihelion passage. It
was very surprising that for both orbital legs the NG~orbits were
perfectly determinable (both NG~solutions are discussed later in
this paper). The O-C~diagrams for pre-perihelion orbits and
post-perihelion orbits are presented in the two upper panels of
Fig~\ref{fig:OC_08a1} for NG~model and pure GR~model, respectively.
One can see the great improvements of NG~orbit fitting for
pre-perihelion data and only slight improvements for post-perihelion
data. One can even get an impression that the NG~orbit determined
from the entire data set (third panel in Fig.~\ref{fig:OC_08a1})
gives a better fitting to post-perihelion data that orbit determined
from the post-perihelion subset of data. Unfortunately, some trends,
in particular in declination, are still noticeable. On the other
hand, however, for NG~orbit based on post-perihelion data we also
noticed the recovery of some measurements taken at the end of
observational arc. Thus, we conclude that NG~orbit based on the
post-perihelion data arc seems to be more adequate to predict the
future of this comet as well as the uncertainty of this prediction.

\begin{savenotes}
\begin{table*}
\caption{\label{tab:models} Characteristics of models taken in this investigation
for previous and next perihelion dynamical evolution. Columns $[3]$
and $[4]$ provide a qualitative assessment of the O-C~distribution
and O-C~diagram, respectively for the types of models given in column [2],
where symbol ++ means distributions well-fitted to a normal distribution
or O-C~diagrams without any meaningful trends in right ascension
or/and declination, + denotes distributions rather similar to a Gaussian
or O-C diagram with slight trends only, - means non-Gaussian distribution
and significant trends in O-C~diagram, - - denotes the worse characteristics
of O-C~distribution or/and O-C~diagram. NG~models with negative
radial component of NG~acceleration, $A_{1}<0$, are marked as NG$_{{\rm A1}}$.
The rms and a qualitative assessment of O-C~distribution are additionally
given in columns $[6]$--$[7]$ for GR~model based on the same interval
of data as NG~model given in column $[2]$. A quality class obtained
in each individual case as well as the $1/{\rm a}_{{\rm ori}}$ and
$1/{\rm a}_{{\rm fut}}$ are given in columns $[8]$--$[10]$, respectively.
In the case of PRE-type of model described in columns $[2]$--$[9]$,
the type of model used for future $1/a$-determination is listed in
column $[11]$.}

\begin{tabular}{@{}llccccccc@{$\pm$}cc@{$\pm$}cl@{}}
\hline
{Comet}  & {Model}  & {Fit to}  & {O-C }  & {rms }  & {rms$_{{\rm GR}}$}  & {GR fit }  & {Orbit}  & \multicolumn{2}{c}{{$1/{\rm a}_{{\rm ori}}$}} & \multicolumn{2}{c}{{$1/{\rm a}_{{\rm fut}}$}} & {Model type for $1/{\rm a}_{{\rm fut}}$} \tabularnewline
{ }  & {type }  & {gauss}  & {}  & {$^{\prime\prime}$}  & {$^{\prime\prime}$}  & {to gauss}  & {class}  & \multicolumn{2}{c}{{$10^{-6}$\,au$^{-1}$}} & \multicolumn{2}{c}{{$10^{-6}$\,au$^{-1}$}} & {if different from $[2]$} \tabularnewline
{$[1]$}  & {$[2]$}  & {$[3]$}  & {$[4]$}  & {$[5]$}  & {$[6]$}  & {$[7]$}  & {$[8]$}  & \multicolumn{2}{c}{{$[9]$}} & \multicolumn{2}{c}{{$[10]$}} & {$[11]$} \tabularnewline
\hline
{2006 HW$_{51}$}  & {GR}            & {+ }  & {+ }  & {0.29}  & {}  & {}  & {1a}  & { 47.31}  & { 3.37}  & { 90.12}  & { 3.37}  & {} \tabularnewline
{}                & {NG$_{\rm A1}$} & {++}  & {++}  & {0.28}  & {0.29}  & {+}  & {1a}  & { 30.32}  & { 4.40}  & { 37.45}  & { 8.34}  & { } \tabularnewline
{}                & {PRE GR}        & {+ }  & {++}  & {0.29}  & {}  & {}  & {2a}  & { 24.84}  & { 6.19}  & \multicolumn{2}{c}{{}} & {} \tabularnewline
\rowcolor{LightGray} 
{2006 K3}         & {NG }           & {++}  & {++}  & {0.54}  & {0.69}  & {+}  & {1a}  & { 61.02}  & \hspace{-3.2mm}{$\pm$\hspace{1.7mm}4.63}  & {$-$131.28}  & \hspace{-3.2mm}{$\pm$\hspace{1.7mm}4.67}  & {} \tabularnewline
{2006 L2}         & {GR }           & {++}  & {+ }  & {0.49}  & {}  & {}  & {1a}  & { 12.89}  & { 1.43}  & { $-$95.28}  & { 1.43}  & {} \tabularnewline
{}                & {NG$_{\rm A1}$} & {+ }  & {++}  & {0.46}  & {0.49}  & {++}  & {1a}  & { 43.56}  & { 4.21}  & {$-$170.88}  & { 7.73}  & { } \tabularnewline
{}                & {PRE GR}        & {++}  & {++}  & {0.37}  & {}  & {}  & {1b}  & { 63.56}  & { 4.52}  & \multicolumn{2}{c}{{}} & {} \tabularnewline
\rowcolor{LightGray} 
{2006 OF$_{2}$}  & {NG }            & {++}  & {+ }  & {0.36}  & {0.38}  & {-}  & {1a+}  & { 21.21}  & \hspace{-3.2mm}{$\pm$\hspace{1.7mm}0.49}  & {$-$658.82}  & \hspace{-3.2mm}{$\pm$\hspace{1.7mm}0.23}  & {} \tabularnewline
{}               & {PRE GR}         & {++}  & {+ }  & {0.36}  & {}  & {}  & {1a}  & { 16.42}  & { 0.62}  & \multicolumn{2}{c}{{}} & {} \tabularnewline
\rowcolor{LightGray} 
{2006 P1}        & {NG }            & {+ }  & {+ }  & {0.25}  & {0.25}  & {+}  & {1b}  & { 57.17}  & \hspace{-3.2mm}{$\pm$\hspace{1.7mm}4.03}  & { 467.65}  & \hspace{-3.2mm}{$\pm$\hspace{1.7mm}3.56}  & {} \tabularnewline
\rowcolor{LightGray} 
{2006 Q1}        & {NG }            & {+ }  & {++}  & {0.37}  & {0.50}  & {- }  & {1a+}  & { 51.08}  & \hspace{-3.2mm}{$\pm$\hspace{1.7mm}0.48}  & { 707.44}  & \hspace{-3.2mm}{$\pm$\hspace{1.7mm}0.31}  & {} \tabularnewline
{}               & {PRE GR}         & {+ }  & {++}  & {0.34}  & {}  & {}  & {1a}  & { 49.44}  & { 0.45}  & \multicolumn{2}{c}{{}} & {} \tabularnewline
{2006 VZ$_{13}$} & {PRE NG}         & {++}  & {+ }  & {0.39}  & {0.40}  & {+ }  & {2a}  & { 13.96}  & { 4.80}  & { 491.21}  & { 20.10}  & {NG, class: 1b}\tabularnewline
{2007 N3}        & {PRE GR}         & {-}   & {+ }  & {0.33}  & {}  & {}  & {1a}  & { 29.31}  & { 0.59}  & { 823.61}  & { 2.06}  & {POST GR~$^{1}$, class: 1a}\tabularnewline
\rowcolor{LightGray}{ }  & {NG }    & {++}  & {+ }  & {0.35}  & {0.50}  & {- }  & {1a+}  & { 32.77}  & \hspace{-3.2mm}{$\pm$\hspace{1.7mm}0.18}  & { 828.64}  & \hspace{-3.2mm}{$\pm$\hspace{1.7mm}0.59}  & {} \tabularnewline
{2007 O1}        & {GR }            & {++}  & {+ }  & {0.47}  & {}  & {}  & {1a}  & { 23.36}  & { 4.70}  & {$-$496.83}  & { 4.69}  & {} \tabularnewline
{2007 Q1}        & {GR }            & {+ }  & {+ }  & {0.58}  & {}  & {}  & {3a}  & { 54.95}  & {799.09}  & {$-$449.88}  & {741.43}  & {} \tabularnewline
{2007 Q3}        & {PRE GR}         & {+ }  & {+ }  & {0.39}  & {}  & {}  & {1a+}  & { 41.91}  & { 0.53}  & { 131.77}  & { 3.63}  & {POST GR, class: 1a}\tabularnewline
\rowcolor{LightGray} { }  & {NG }   & {+ }  & {- }  & {0.39}  & {0.49}  & {- }  & {1a+}  & { 39.13}  & \hspace{-3.2mm}{$\pm$\hspace{1.7mm}0.49}  & { 118.96}  & \hspace{-3.2mm}{$\pm$\hspace{1.7mm}0.96}  & {} \tabularnewline
{2007 W1}        & {PRE NG}         & {+ }  & {+ }  & {0.49}  & {0.61}  & {- -}  & {1b}  & {$-$42.71~~}  & { 2.34}  & { 554.38}  & { 7.09}  & {POST NG, class: 2a} \tabularnewline
\rowcolor{LightGray} {} & {NG$+\tau$~~$^{2}$ } & {- }  & {- }  & {0.67}  & {2.96}  & {- -}  & {1a}  & {$-$36.56~~}  & \hspace{-3.2mm}{$\pm$\hspace{1.7mm}1.86}  & { 549.97}  & \hspace{-3.2mm}{$\pm$\hspace{1.7mm}5.79}  & {} \tabularnewline
\rowcolor{LightGray} 
{2007 W3}  & {NG }  & {++}  & {+ }  & {0.52}  & {0.54}  & {++}  & {1b}  & { 31.38}  & \hspace{-3.2mm}{$\pm$\hspace{1.7mm}3.85}  & { 343.89}  & \hspace{-3mm}{$\pm$\hspace{0.4mm}18.10}  & {} \tabularnewline
{2008 A1}  & {PRE NG}  & {++}  & {++}  & {0.28}  & {0.47}  & {}  & {1b}  & {120.84}  & { 2.03}  & { 246.52}  & { 2.82}  & {POST NG, class: 1b}\tabularnewline
\rowcolor{LightGray} {}  & {NG$+\tau$~~$^{2}$}  & {+ }  & {- }  & {0.44}  & {1.44}  & {- -}  & {1a}  & {123.07}  & \hspace{-3mm}{$\pm$\hspace{0.4mm}1.50}  & { 256.41}  & \hspace{-3mm}{$\pm$\hspace{0.4mm}2.24}  & {} \tabularnewline
{}  & {NG }  & {+ }  & {- }  & {0.45}  & {1.44}  & {- -}  & {1a}  & {120.14}  & { 1.56}  & { 247.21}  & { 1.89}  & {} \tabularnewline
{2008 C1}  & {GR }  & {++}  & {++}  & {0.36}  & {}  & {}  & {2a}  & { 38.57}  & { 11.77}  & { 502.56}  & { 11.77}  & {} \tabularnewline
{}  & {NG }  & {++}  & {++}  & {0.36}  & {0.36}  & {++}  & {2a}  & {115.95}  & { 59.16}  & { 648.87}  & {221.41}  & {} \tabularnewline
{2008 J6}  & {GR }  & {++}  & {+ }  & {0.47}  & {}  & {}  & {1b}  & { 25.35}  & { 4.00}  & {$-$479.69}  & { 3.99}  & {} \tabularnewline
{2008 T2}  & {GR }  & {++}  & {+ }  & {0.38}  & {}  & {}  & {1b}  & { 12.22}  & { 1.06}  & { 275.92}  & { 1.06}  & {} \tabularnewline
{}  & {NG$_{{\rm A1}}$}  & {++}  & {++}  & {0.39}  & {0.38}  & {++}  & {1b}  & { 19.19}  & { 1.14}  & { 218.90}  & { 7.83}  & { } \tabularnewline
{}  & {PRE GR}  & {++}  & {++}  & {0.39}  & {}  & {}  & {1b}  & { 11.47}  & { 1.08}  & \multicolumn{2}{c}{{}} & {} \tabularnewline
{2009 K5}  & {PRE GR}  & {- }  & {+ }  & {0.33}  & {}  & {}  & {1a}  & { 45.50}  & { 0.55}  & { 552.91}  & { 0.41}  & {POST GR, class: 1a}\tabularnewline
{}  & {GR }  & {- }  & {- }  & {0.47}  & {}  & {- -}  & 1a+  & { 49.42}  & { 0.22}  & { 554.68}  & { 0.22}  & {} \tabularnewline
{}  & {DIST NG$_{{\rm A1}}$}  & {+ }  & {++}  & {0.39}  & {0.38}  & {- -}  & {1a}  & { 44.22}  & { 2.75}  & { 550.33}  & { 1.60}  & { } \tabularnewline
{2009 O4}  & {GR }  & {+ }  & {++}  & {0.39}  & {}  & {}  & {1b}  & { 55.96}  & { 4.91}  & { $-$55.96}  & { 4.91}  & {} \tabularnewline
\rowcolor{LightGray} 
{2009 R1}  & {NG }  & {++}  & {++}  & {0.51}  & {0.63}  & {+ }  & {1b}  & { 12.16}  & \hspace{-3.2mm}{$\pm$\hspace{1.7mm}3.29}  & { 170.43}  & \hspace{-2mm}{$\pm$\hspace{0.2mm}197.25}  & {} \tabularnewline
{2010 H1}  & {GR }  & {+ }  & {+ }  & {0.83}  & {}  & {}  & 2b  & {240.15}  & { 75.41}  & { 784.70}  & { 75.46}  & {} \tabularnewline
{2010 X1}  & {GR$_{{\rm May}}$}  & {+ }  & {+ }  & {0.46}  & {}  & {}  & {1b}  & { 24.10}  & { 2.20}  & \multicolumn{2}{c}{disintegration} & {} \tabularnewline
{}  & {GR$_{{\rm Apr}}$}  & {++}  & {++}  & {0.37}  & {}  & {}  & {1b}  & { 27.24}  & { 3.95}  & \multicolumn{2}{c}{} & {} \tabularnewline
\hline
\end{tabular}
{\footnotesize \hspace{2.5cm}$^{1}$~~Model based on data started
from heliocentric distance of 2.0\,au from the Sun }\\
{\footnotesize ~~~~$^{2}$~~Asymmetric models relative to perihelion passage based on entire data sets}
\end{table*}

\end{savenotes}

The most manifesting NG~accelerations we found in C/2007~W1 case
where a similar approach was necessary to be used as for C/2008~A1.
The same method was applied earlier for C/2007~W1 by Nakano
\citeyearpar{NK1731A,NK1731B}. We found that the orbit obtained on
the basis of the whole data set proved to be inadequate to describe
the actual motion of both comets (C/2007~W1 and C/2008~A1). We
noticed this fact by subscript 'un' (i.e. uncertain) in column
{[}10{]} of Table~\ref{tab:Obs-mat}. Among comets with very long
data intervals ($>$~3\,yr) there are two other objects (C/2007~N3
and C/2007~Q3) with well visible trends in O-C~diagrams taken for
NG~orbit and based on the full observational interval. Therefore,
for these two comets it was also necessary to determine the
osculating orbit from pre- and post-perihelion subsets of data,
independently. Further, in the comet C/2006~VZ$_{13}$ case, an
adequate NG~orbit for past dynamical evolution were determined only
for some part of pre-perihelion data (ending 40~days before
perihelion passage).

Analysing orbits of eleven comets with well-detectable NG~effects
in their orbital motion we concluded that the motion of six of them
can be model by one set of NG~parameters during the period covered
by data. In the case of remaining five comets the NG~behaviour is
more complicated. Thus, we proposed in these cases to model their
motion separately before and after perihelion passage.

It turn out that NG~effects are easily visible in the time interval
covered by positional observations also in the motion of comet C/2010~X1.
However, this is very special case because the nature of this NG~behaviour
is likely to be violent -- this comet started to disintegrate at least
about one months before perihelion passage (in August 2011). Therefore,
the NG~model used here is completely inadequate (see also Section~\ref{sub_cometary_case_B}).
Thus, we modelled the motion of this comet based on the shorter interval
of data, e.g. carried out before the disintegration process was started.
It turn out, that the best orbit (GR~orbit), for the past evolution
was determined from an arc of data limited to the observations taken
before 1~May 2011 what can suggest that a disruption started even
before August~2011. For this reason we also marked the GR~solution
based on almost entire data interval by subscript 'un' in Table~\ref{tab:Obs-mat}.

For other three comets, C/2006~HW$_{51}$, C/2006~L2 and C/2008~T2,
the NG~effects are worse detectable than in previous cases. Additionally,
the radial component of standard NG~acceleration, A$_{1}$ (see Eq.~\ref{eq:ng_std}),
is negative for these objects (see also Section~\ref{sub_NGmotion}).
However, the GR~orbit is quite acceptable since only slight trends
are visible in the O-C diagrams of these comets. Therefore, we discussed
for them the GR~orbit (Table~\ref{tab:Obs-mat}) as at least similarly
likely solution as NG~orbit (Table~\ref{tab:models}) and included
to the group of comets with weak NG~effects (Section~\ref{sub_cometary_case_C}).

Comet C/2009~K5 also have marginally detectable NG~effects with
a negative A$_{1}$ but it differs from the previous three comets
in visible trends in the O-C~diagrams for GR~orbit as well NG~orbit.
Thus, we included this comet to special objects (marked by subscript
'un' in column $[10]$ of Table~\ref{tab:Obs-mat}) where it was
necessary to divide the data into pre-perihelion and post-perihelion
orbital branches (see Section~\ref{sub_cometary_case_B}).

An interesting example is comet C/2008~C1 that has marginally detectable
NG~effects. This is surprising because of short interval of data
(0.3 yr). We decided to use the GR~solution as more reliable for
this comet of second quality orbit (Table~\ref{tab:Obs-mat}).

NG~effects are completely not detectable in the orbital motion of
five comets, C/2007~O1, C/2007~Q1, C/2008~J6, C/2009~O4, C/2010~H1,
and for these objects we have just pure GR osculating orbits (Table~\ref{tab:Obs-mat}).
All these comets have osculating perihelion distance ${\rm q}_{{\rm osc}}>2$\,au
and two of them have poor quality orbits (see Table~\ref{tab:Obs-mat}).

\subsection{Osculating orbit determination from some part of data}

\label{sub_orbit_part_data}

According to the brief overview of orbital determination given in
Section~\ref{sub_orbit_full_data} 
we have seven special comets, C/2006~VZ$_{13}$, C/2007~N3, C/2007~Q3,
C/2007~W1, C/2008~A1, C/2009~K5 and C/2010~X1, where we have to
apply a non-standard approach and determine two individual osculating
orbits (GR~or NG~if possible) for two orbital legs. These solutions
were marked as PRE and POST models in column $[2]$ of Table~\ref{tab:models}.
For C/2009~K5 we additionally presented DIST model based on data
taken from the heliocentric distances larger than $r_{h}>2.5$\,au.

For the remaining comets we noticed that osculating orbits (GR or
NG, see Table~\ref{tab:Obs-mat}) based on the entire data interval
well represent their orbital motion in the period covered by observations.
However, for some of these comets (wherever we though it relevant
and meaningful) we also determined the osculating orbits based on
only a subset of pre-perihelion measurements (i.e. for C/2006~OF$_{2}$,
C/2006~Q1 and C/2008~C1). These additional orbital solutions are
presented here just for comparison and wider discussion of the dynamical
evolution of each comet carried out in Part~II. Moreover, as was
discussed in Section~\ref{sub_orbit_full_data}, for comets C/2006~HW$_{51}$,
C/2006~L2 and C/2008~T2 the NG~models based on entire set of data
(where radial component of NG~acceleration is negative) are also
given just for comparison.

Generally, in Table~\ref{tab:models} are presented all models (of
osculating orbit) taken further in the second part of this
investigation for previous and next perihelion dynamical evolution.
The best model derived for each comet is given always in the first
row and just these best models are used in
Section~\ref{sec_orbit_original_future} for statistical analysis.
The NG~models based on the entire data sets are shown in coloured
light grey rows, similarly as in Table~\ref{tab:Obs-mat}.

\section{Accuracy of the cometary orbit}

\label{sec_orbit_accuracy}

\begin{table}
\caption{\label{tab:mse78_II} Quantities for establishing accuracy of orbit. 
Slightly modified new version of Table~II from MSE}

\centering{}%
\begin{tabular}{@{}rr@{~,~}lr@{~,~}l}
\hline
$L$ \& $M$  & \multicolumn{2}{c}{{Mean error of 1/a$_{{\rm osc}}$}} & \multicolumn{2}{c}{{time span of observations}}\tabularnewline
 & \multicolumn{2}{c}{{in units of $10^{-6}$\,au$^{-1}$}} & \multicolumn{2}{c}{{in months or days}}\tabularnewline
\hline
{8}  & \multicolumn{2}{c}{{}} & \multicolumn{2}{l}{{~~~~$\geq$ 48 months}}\tabularnewline
{7}  & \multicolumn{2}{l}{{~~~~~~~~$<$ 1}} & {$[$24}  & {48$[$} \tabularnewline
{6}  & {$[$1}  & {5$[$}  & {$[$12}  & {24$[$} \tabularnewline
{5}  & {$[$5}  & {20$[$}  & {$[$ 6}  & {12$[$} \tabularnewline
{4}  & {$[$20}  & {100$[$}  & {$[$ 3}  & {6$[$} \tabularnewline
{3}  & {$[$100}  & {500$[$}  & {$[$1.5}  & {3$[$} \tabularnewline
{2}  & {$[$500}  & {2500$[$}  & {$[$23} days  & {1.5 months$[$} \tabularnewline
{1}  & {$[$2\,500}  & {12\,500$[$}  & {$[$12}  & {23$[$ days} \tabularnewline
{0}  & \multicolumn{2}{c}{{$\geq$ 12\,500}} & {$[$7}  & {12$[$} \tabularnewline
{-1}  & \multicolumn{2}{c}{{}} & {$[$3}  & {7$[$} \tabularnewline
{-2}  & \multicolumn{2}{c}{{}} & {$[$1}  & {3$[$} \tabularnewline
\hline
\end{tabular}
\end{table}

In 1978 MSE formulated the recipe to evaluate the accuracy of the
osculating cometary orbits obtained from the positional data. They
proposed to measure this accuracy by the quantity $Q$ defined as

\begin{equation}
Q=1/2\cdot(L+M+N)+\delta,{\rm ~~~where}\label{eq:orbit_accuracy}
\end{equation}

\noindent $L$ denotes a small integer number which depends on the
mean error of the determination of the osculating 1/a,

\noindent $M$ -- a small integer number which depends on the time
interval covered by the observations,

\noindent $N$ -- a small integer number that reflects the number
of planets, whose perturbation were taken into account, and

\noindent $\delta$ equals 0.5 or 1 to make $Q$ an integer number.

\vspace{0.1cm}

\noindent Values of $L$, $M$ and $N$ are obtained following the
scheme presented in their original Table~II (MSE). 
According to MSE the integer $Q$-value calculated from Eq.~\ref{eq:orbit_accuracy}
should be next replaced with the orbit quality class as follows: value
of $Q=9,8$ means orbit of 1A orbital class, $Q=7$ -- 1B class, $Q=6$
-- 2A class, $Q=5$ -- 2B class, and $Q<5$ means the worse than second
class orbit.

\vspace{0.1cm}

There are three reasons for which we found that some slight, but important adjustment 
of the above orbital accuracy assessment should be done:

\begin{enumerate}
\item In the modern orbit determination all Solar system planets are taken into account, 
therefore we always have $N=3$. Thus, Eq.~\ref{eq:orbit_accuracy} can be
rewritten for our purpose in a form:
\begin{eqnarray}
 &  & Q=Q^{*}+\delta,{\rm ~~where~~}\nonumber \\
 &  & Q^{*}=0.5\cdot(L+M+3){\rm ~~~~and~~~~}\delta=1{\rm ~~~or~~~}0.5\label{eq:orbit_accuracy_2}
\end{eqnarray}
\item Current  cometary observations are generally of a high precision and the resulting osculating 1/a-uncertainties 
are often smaller than 10$^{-6}$\,au$^{-1}$. It should be reflected in a higher $L$ number than is allowed 
in the original MSE table. Similarly, a higher $M$ number should be attributed to very long time intervals covered by
contemporary positional observations of LPCs. Thus, the mean error of $1/a_{\rm osc}$ smaller than 1 unit (i.e.
$1\times10^{-6}$\,au$^{-1}$) gives now $L=7$ and a time span of
data can be greater than 48~months ($M=8$, see Table~\ref{tab:mse78_II}). 
\item Almost all orbits of currently discovered LPCs would be classified as the 1A quality class. 
For example, we found that using the original $\delta$~definition (to be 0.5 or 1) we obtained 15~comets of orbital
class 1A and just two comets (C/2007~Q1 and C/2010~H1) of an obvious cases of second
or worse orbital class (extremely short orbital arcs) for the sample studied here. 
Thus, additional modification seems to be necessary to obtain 
a better diversification of orbit accuracy classes. We proposed below, the new $\delta$ 
definition and division the first quality class into three new classes (1a+, 1a, 1b) instead former 1A and 1B. 

\end{enumerate}

Thus, the more relevant definition of $Q$ for currently discovered LPCs takes the form:
\begin{eqnarray}
 &  & Q=Q^{*}+\delta,{\rm ~~where~~}\nonumber \\
 &  & Q^{*}=0.5\cdot(L+M+3){\rm ~~~~and~~~~}\delta=0{\rm ~~~or~~~}0.5\label{eq:orbit_accuracy_3}
\end{eqnarray}
How to calculate the quantities $L$ and $M$ is described in Table~\ref{tab:mse78_II}
that is a simpler form of original Table~II given by MSE. 
Value of $\delta$ equals 0.5 or 0 to make $Q$ an integer number.
When $Q^{*}$ is an intermediate between two integers (that define
two consecutive orbital classes) the proposed new recipe gives the final $Q$-values exactly
the same as originally proposed by MSE.

To distinguish the proposed quality system from MSE~system,
we use lower-case letter 'a' and 'b' in quality class descriptions
instead of original 'A' and 'B' in the following way: $Q=9$ -- 1a+
class, $Q=8$ -- 1a class, $Q=7$ -- 1b class, $Q=6$ -- 2a class, $Q=5$ -- 2b class,
$Q=4$ -- 3a class, $Q=3$ -- 3b class, and $Q\le2$ -- 4 class, where
$Q$ is calculated according to eq.~\ref{eq:orbit_accuracy_3}. We
additionally propose to introduce a special 1a+ quality class in case
of $Q=9$. The quality classes 3a, 3b and 4 were not defined by MSE,
but we adopted here the idea published by Minor Planet Center (at
http://www.minorplanetcenter.net/iau/info/UValue.html) as 'a logical
extension to the MSE~scheme'.

This new orbit quality scheme separates the orbits of a very good
quality in MSE~system, 1A, into three quality classes in the new
system, where the worst of orbits in 1A class (Q$^{*}$=7) in MSE
are classified as 1b in the new scheme. The reader can check how it
works using the Q$^{*}$ value given in last column of Table~\ref{tab:Obs-mat}
to calculate the MSE quality class.

In the MWC\,08 only pure GR~orbits are classified according
to MSE~method. It seems, however, that there are no contraindications
to use such a procedure to qualify also NG~orbits. These NG~orbits
are often of a lower quality than GR~orbit obtained on the basis of
the same time interval. This is an obvious consequence of higher uncertainties
of NG~orbital elements as a result of additional NG~parameters to
determine simultaneously with six orbital elements. In our opinion,
the quality comparison between GR~orbit and NG~orbit should not
be carried out exclusively on the basis of an orbital quality because
NG~orbit is always closer to reality by its physical assumptions,
and therefore is more appropriate to describe the cometary motion
than GR~orbit. On the other hand, a quality assessment for NG~orbit
gives an easy indication of the uncertainty of orbital elements.

It is worth to stress, that any effective orbit quality assessment
system (including the MSE method and this proposed here) describes a
particular orbital solution based on a given data set (or subset),
procedure used for this data processing as well as the adopted force
model. Therefore, orbit of each individual comet can be classified
differently, depending on the preferred orbital solution.

\subsection{Application of the modified method of orbital quality assessment
to comets considered so far}
\label{sub_orbit_accuracy_application}

The new modified MSE~procedure described above, was applied for the determination
of orbital classes of all osculating orbits of the investigated here comets.
The results are given in columns~12 and 9 of Table~\ref{tab:Obs-mat}
and in column~8 of Table~\ref{tab:models}. As was mentioned above,
one can notice the higher diversification in orbital classes between
investigated comets than using the original MSE~recipe. This can
easily be checked using Eq.~\ref{eq:orbit_accuracy_2} and values
given in column 12 of Table~\ref{tab:Obs-mat}. According to the
proposed method we have a lower fraction of comets of the best orbital
classes: 11 (1a+ and 1a classes) instead of 15 (1A class). 
Additionally, these 11 comets are now divided 
into 5 comets of 1a+ class and 6 comets of 1a~class when using entire
data sets shown in Table~\ref{tab:Obs-mat}. Moreover, the orbit
of one comet, C/2008~C1, is reclassified as a second class orbit.

We noticed that at the IAU~Minor Planet Center web pages orbital quality codes are given
also for newly discovered comets. In February 2013, for five comets
investigated in this paper, the quality codes are not shown, namely
for C/2007~Q1 (extremely short arc of data) C/2007~W1 (NG orbit),
C/2008~A1 (NG orbit), C/2008~T2 (NG orbit with negative radial component
of NG~acceleration) and C/2010~X1. The remaining 17~objects are
classified as follows: 13 comets have 1A class, 2 comets -- class
1B (C/2006~VZ$_{13}$, C/2010~H1), and 2 objects -- of class 2A
(C/2008~C1, C/2009~O4). We also estimated the orbit of C/2008~C1
as a 2a quality class, but orbit of C/2009~O4 as 1b class. 
However, 2A~quality code for C/2009~O4 is there a natural consequence of shorter
intervals of data taken for orbit determination than in the present investigation. 
Similarly, the \citet{JPL_Browser} shows (in February
2013) the osculating orbits on the basis of full interval of positional
data.

\begin{table}
\caption{\label{tab:orbital_quality86_list} New quality class assessment for
86~near-parabolic comets investigated in Paper1\&2}

\centering{}{\setlength{\tabcolsep}{2.0pt} %
\begin{tabular}{llllllll}
\hline
~~C/  & Class  & ~~C/  & Class  & ~~C/  & Class  & ~~C/  & Class \tabularnewline
\hline
 &  &  &  &  &  &  & \tabularnewline
\multicolumn{8}{c}{37 comets of NG orbits }\tabularnewline
 &  &  &  &  &  &  & \tabularnewline
\multicolumn{8}{c}{$q_{{\rm osc}}<3.1$\,au}\tabularnewline
1885 X1  & 2a  & 1892 Q1  & 2a  & 1913 Y1  & 1b  & 1940 R2  & 2a \tabularnewline
1946 U1  & 1b  & 1952 W1  & 2a  & 1956 R1  & 1b  & 1959 Y1  & 2b \tabularnewline
1974 F1  & 1a  & 1978 H1  & 1b  & 1986 P1  & 1a  & 1989 Q1  & 2a \tabularnewline
1989 X1  & 1b  & 1990 K1  & 1a  & 1991 F2  & 1b  & 1993 A1  & 1a \tabularnewline
1993 Q1  & 1b  & 1996 E1  & 1b  & 1997 J2  & 1a+ & 1999 Y1   & 1a+ \tabularnewline
2001 Q4  & 1a+  & 2002 E2  & 1b  & 2002 T7  & 1a+  & 2003 K4  & 1a+ \tabularnewline
2004 B1  & 1a+  &  &  &  &  &  & \tabularnewline
\multicolumn{8}{c}{$q_{{\rm osc}}\ge3.1$\,au}\tabularnewline
1980 E1  & 1a+  & 1983 O1  & 1a  & 1984 W2  & 1a  & 1997 BA6  & 1a+ \tabularnewline
1999 H3  & 1a  & 2000 CT$_{54}$  & 1a+  & 2000 SV$_{74}$  & 1a+  & 2002 R3  & 1a \tabularnewline
2005 B1  & 1a+  & 2005 EL$_{173}$  & 1a+  & 2005 K1  & 1a  & 2006 S2  & 1b \tabularnewline
 &  &  &  &  &  &  & \tabularnewline
\multicolumn{8}{c}{49 comets of GR orbits }\tabularnewline
 &  &  &  &  &  &  & \tabularnewline
\multicolumn{8}{c}{$q_{{\rm osc}}<3.1$\,au}\tabularnewline
1992 J1  & 1a+  & 2001 K3  & 2a  &  &  &  & \tabularnewline
\multicolumn{8}{c}{$q_{{\rm osc}}\ge3.1$\,au}\tabularnewline
1972 L1  & 1a  & 1973 W1  & 1b  & 1974 V1  & 1b  & 1976 D2  & 1a \tabularnewline
1976 U1  & 1b  & 1978 A1  & 1a  & 1978 G2  & 1b  & 1979 M3  & 1b \tabularnewline
1987 F1  & 1a  & 1987 H1  & 1a+  & 1987 W3  & 1a  & 1988 B1  & 1a \tabularnewline
1993 F1  & 1a  & 1993 K1  & 1a  & 1997 A1  & 1b  & 1999 F1  & 1a+ \tabularnewline
1999 F2  & 1a  & 1999 J2  & 1a+  & 1999 K5  & 1a+  & 1999 N4  & 1a \tabularnewline
1999 S2  & 1a  & 1999 U1  & 1a  & 1999 U4  & 1a+  & 2000 A1  & 1a \tabularnewline
2000 K1  & 1a  & 2000 O1  & 1a  & 2000 Y1  & 1a  & 2001 C1  & 1a \tabularnewline
2001 G1  & 1a  & 2001 K5  & 1a+  & 2002 A3  & 1a  & 2002 J4  & 1a \tabularnewline
2002 J5  & 1a+  & 2002 L9  & 1a+  & 2003 G1  & 1a  & 2003 S3  & 1a \tabularnewline
2003 WT$_{42}$  & 1a+  & 2004 P1  & 1a  & 2004 T3  & 1a  & 2004 X3  & 1a \tabularnewline
2005 G1  & 1a  & 2005 Q1  & 1a  & 2006 E1  & 1a  & 2006 K1  & 1a+ \tabularnewline
2006 YC  & 2a  & 2007 JA$_{21}$  & 1a  & 2007 Y1  & 2a  &  & \tabularnewline
\hline
\end{tabular}}
\end{table}

We also applied the modified method of orbital quality determination
to pure GR as well as NG~orbits of comets from Papers~1 \& 2, where all those Oort spike
comets were chosen from MWC\,08 as objects with highest quality orbits
(classes 1A or 1B). The full list of these new quality class assessment 
for the entire sample of near-parabolic comets is given in
Table~\ref{tab:orbital_quality86_list}. An extended table, for
108~LPCs investigated by us that includes details about the
observational material used for orbit determination, type of models
used, as well as the orbit quality estimation, is available in our
web page \citep{dyb-krol-web:2013}.

It turned out that among 37~comets with NG~orbits,
six (C/1885~X1, C/1892~Q1, C/1940~R2, C/1952~W1, C/1959~Y1 and C/1989~Q1) 
should be classified as second quality orbit according to a modified,
more restrictive method proposed here (Table~\ref{tab:orbital_quality86_list}).
Next three with pure GR orbit are now also class 2 objects 
(C/2001~K3, C/2006~YC and C/2007~Y1).

Summarizing, according to a more restrictive orbital quality
assessment proposed here we obtained 23 comets of 1a+~class, 38
comets of 1a~class, 16 -- 1b~class, 8 -- 2a~class, and 1~object of
2b~class in the sample of 86~comets analysed in Papers~1 \& 2 that
were chosen from MWC\,08 as Oort spike comets having pure
GR orbit of the highest quality (class 1) or NG~orbit (then
the quality class is not specified in the catalogue).

\section{Implication of data structure on orbit determination}
\label{sec_cometary_cases}

In this section we show how the data structure influence the proper
choice of the tailored orbit determination method, how it affects
the quality of osculating orbits and the possibility of the determination
of NG~effects in the motion of near-parabolic comet. We
performed an extensive review and tests of various models for each
of investigated comets, including asymmetric ones 
(i.e. NG~models with $\tau$-shift of maximum of $g(r)$-function relative to perihelion passage, 
see section~\ref{sub_NGmotion}) and orbital modelling based only 
on one branch of their orbits (pre- or post-perihelion, see section~\ref{sub_orbit_part_data}).
From the entire range of solutions obtained for these
comets we included here only the best models, in the sense of three
criteria mentioned in Section~\ref{sub_orbit_full_data}: the decrease
of rms, the similarity of O-C~distribution to the Gaussian distribution
and the absence of trends in O-C~diagrams, and keeping a minimum
number of necessary NG~parameters in the case of NG~models (Table~\ref{tab:models}). 
We found useful, for this aspects of further discussion, to divide
the investigated sample of comets into four groups (see also column
11 of Table~\ref{tab:Obs-mat}): 
\begin{description}
\item [group A:] four comets with more than 3 yrs of data and heliocentric
distance span at least from 6\,au before perihelion to 6\,au after
perihelion, 
\item [group B:] eight peculiar comets, e.g. extremely bright or
and/or with detected split event and/or strong/variable NG~effects
in their motion, 
\item [group C:] six comets of non-detectable or very weak NG~effects
in their motions, 
\item [group D:] four comets of a second or worse
quality of osculating orbit. 
\end{description}

Below we discuss the differences in data sets between these groups
and whether and how these differences affect the precise determination
of NG~orbit. The reference table to individual solutions discussed in this section
is implicitly Table~\ref{tab:models}.

\subsection{Group A: comets with long time-intervals of observations}

\label{sub_cometary_case_A}

\begin{figure}
\includegraphics[width=8.8cm]{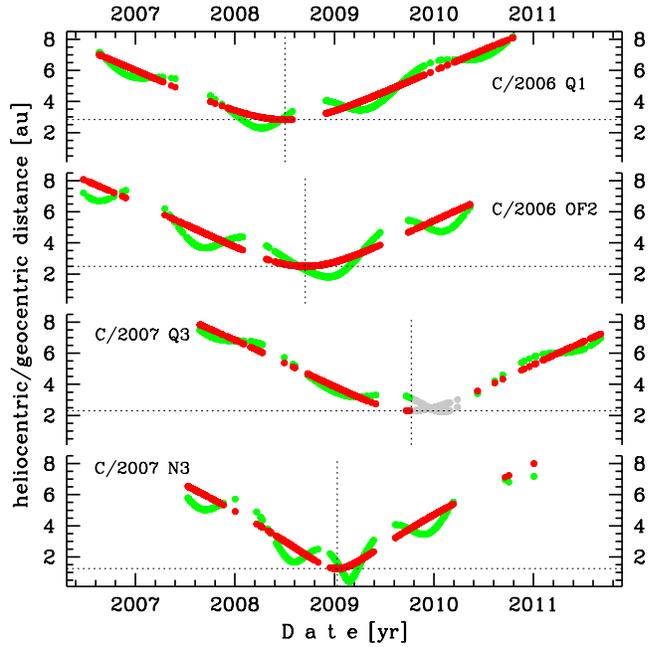} 
\caption{Comets with long time intervals of data arranged in order of decreasing
$q_{\rm osc}$ (from top to bottom). Time distribution of positional
observations with corresponding heliocentric (red curve) and geocentric
(green curve) distance at which they were taken. Horizontal dotted
lines show the perihelion distance for a given comet whereas vertical
dotted lines -- the moment of perihelion passage. The horizontal axis
for all panels is exactly the same. For Comet C/2007~Q3 time interval,
where the O-C~diagrams give significant trends in right ascension
and declination (remains visible even in a NG~model) is shown in
light gray -- this period correlates with the moment when a secondary
fragment broke away from the nucleus.}
\label{fig:DataDistLong} 
\end{figure}

One can see in Fig.~\ref{fig:DataDistLong} that all
four comets of this group have been observed more than three years
in a broad range of heliocentric distances from at least 6\,au before
the perihelion passage to over 6\,au after perihelion. Thus, long
time sequences of data should allow us to model the NG~orbital motion
in great details. In particular, these should allow for examining
various forms of $g(r)$-like function (Paper~3). 

In fact, long time series of data allow us to determine
the NG~effects of all these comets basing on the entire ranges of
data, and all their NG~orbits are of the highest accuracy (1a+ class,
see Table~\ref{tab:Obs-mat}). In all cases where we were able to
apply an asymmetric NG~model relative to the moment of perihelion
passage (Section~\ref{sub_NGmotion}), such a model however
did not show any decrease in rms in comparison to a symmetric NG~model
(Table~\ref{tab:NG-parameters})
and did not give a better similarity of O-C~distribution to a normal
distribution and/or better O-C~diagram. Moreover,
it was not possible to derive any dedicated form of $g(r)$-like functions
for these comets. We were also not able to formulate any concrete
conclusions about the potential deviation of value of the exponent
$m$ in Eq.~\ref{eq_NG_model:GEN} from the standard value $m_{\rm STD}=-2.15$,
or about the different value of scale distance than the standard $r_{0}=2.808$\,au
(Eq.~\ref{eq:ng_std}).

\subsubsection*{C/2006~OF$_{2}$~Broughton}

NG~models of this comet provide O-C~distributions very good approximated by Gaussian distribution 
and give very reasonable values of
NG~parameters with dominant and positive radial component of NG~acceleration 
and negligible normal component in comparison to the remaining two $A_1$ and $A_2$ 
components (Table~\ref{tab:NG-parameters}). 
However, even assuming $A_3=0$, the tau-shift can not be determined 
(its value oscillates with large amplitude around more than 100 days before perihelion). 
Due to some slight trends in the O-C~diagram
we added the gravitational PRE model as an alternative for studying past dynamics of this object.

\subsubsection*{C/2006~Q1~McNaught}

Here, NG~models result in O-C~distributions good approximated
by Gaussian distribution and no trends in O-C-diagram was noticed.
The normal component of NG~motion in the standard MSY model seems to be important in orbital fitting and 
gives a significantly decrease of rms in comparison to model with two NG~parameters 
(radial and transverse components) as well as in model 
including also $\tau$-shift of $g(r)$-function and ignoring normal component (Table~\ref{tab:NG-parameters}). 
We added the gravitational PRE model as an alternative for studying past dynamics of this object -- just for comparison.

\subsubsection*{C/2007~N3~Lulin}

This comet has the smallest perihelion distance in this group, $q_{\rm osc}=1.21$\,au.
Therefore, starting this investigation we suspected that we would
get some interesting information about $g(r)$-like function for this
comet. Unfortunately, models based on individually adjusted $g(r)$-like
function did not give noticeably better fitting to observations. 

For NG~model with three components of NG~accelerations the significant decrease of rms was noticed from 0\farcs 50 to 0\farcs35. It can be seen (Table~\ref{tab:NG-parameters}) that values of three NG~parameters, $A_1, A_2, A_3$, although small, are well-defined, however A3 component slightly dominates over $A_1$ and $A_2$. Asymmetric model with two components of NG accelerations and tau ($A_1+A_2+\tau$) gives the rms of 0\farcs49 (significantly greater than symmetric model with $A_3$) Asymmetric model with four parameters ($A_1+A_2+A_3+\tau$) gives rms of 0\farcs35, thus indistinguishable from the symmetric model with three parameters ($A_1+A_2+A_3$, see Table 2).

Moreover, all considered NG~models based on the entire interval of observations give the O-C~distribution that substantially differs from a normal distribution.
For that reason we would recommend gravitational models PRE and POST
instead, especially for studying past and future dynamics
of this object. 

NG~solution based on entire data set as well PRE (POST) type of solution give very similar values of $1/a_{\rm ori}$ ($1/a_{\rm fut}$). Thus, past and future dynamics of this comet seems to be very well defined.

\subsubsection*{C/2007~Q3~Siding~Spring }

This comet must be examined in a special way. In the middle of March
2010, Nick Howes reported a small secondary piece of C/2007~Q3 on
the picture taken using Faulkes Telescope North. The existence of
this secondary component was later confirmed during the follow-up
observations taken from Mar.~17 up to Apr.~9 by many other observers
\citep{colas-man-how-bry:2010}. Indeed, this period correlates with
a time interval where significant trends in both right ascension and
in declination appear in the O-C diagrams even in the NG~model of
motion (light grey part of data in the third panel from the top of
Fig.~\ref{fig:DataDistLong}). GR~orbit of C/2007~Q3
based only on pre-perihelion data is still 1a+~quality class.

Maybe due to this additional event, the NG model of C/2007~Q3 based on entire data interval
gives radial component, $A_1$, smaller than the other two components $A_2$ and $A_3$ (Table~\ref{tab:NG-parameters}).
Similarly as in C/2007~N3 we noticed here important role of a normal component of NG~acceleration in rms decreasing.

Similarly as in remaining comets in this group, the NG~solution based on entire data set as well PRE (POST) type of solution give similar values of $1/a_{\rm ori}$ ($1/a_{\rm fut}$). Thus, past and future dynamics also for this comet seems be well-defined.

\subsection{Group B: peculiar comets}
\label{sub_cometary_case_B}

\begin{figure}
\includegraphics[width=8.8cm]{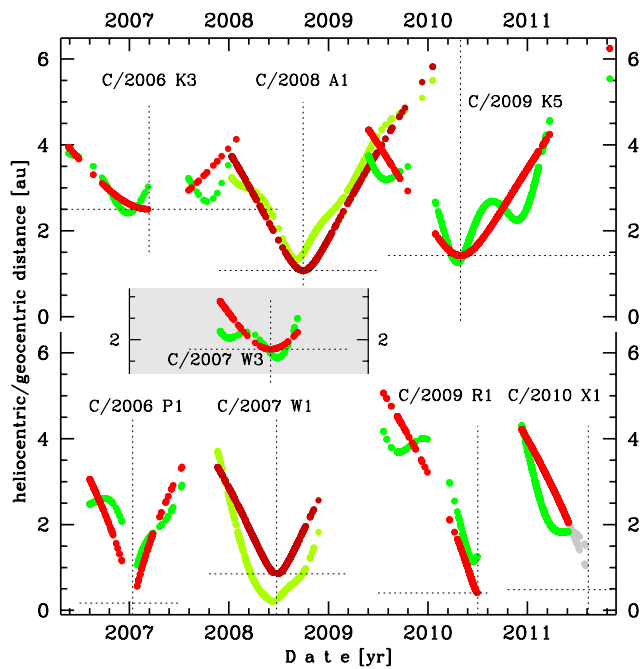} 
\caption{The same as in Fig.~\ref{fig:DataDistLong} for peculiar comets.
The horizontal axis is common for all comets and is the same as in
Fig.~~\ref{fig:DataDistLong}. The vertical scale unit is also the
same as in the previous figure. Four comets with perihelion distances,
$q_{\rm osc}$, below 1.0\,au are shown in the bottom part of this
figure.}
\label{fig:DataDistPeculiar} 
\end{figure}

We found that eight comets in the studied sample exhibit some kind
of unusual activity or are troublesome in determining the osculating
orbit. All of them we classified as peculiar objects and
discussed in this section.

In the bottom part of Fig.~\ref{fig:DataDistPeculiar} the positional
measurements for four comets with the smallest perihelion distances
in our sample are shown: C/2006~P1, C/2009~R1, C/2010~X1 and C/2007~W1. 

Among the remaining four comets in this group we have C/2008~A1 with
strong and variable NG~effects in its motion, two more comets with NG~effects clearly seen in the motion
and easily determinable from the entire interval of data (C/2007~W3
and C/2006~K3), and C/2009~K5, whose osculating orbit
was especially difficult to determine (see below).

\begin{table*}
\caption{\label{tab:NGTwoComets} NG~parameters derived in the present investigation
and by Nakano \citep{NK1731A,NK1731B,NK1807} for two comets with
strongly manifesting NG~effects in the motion.}

\centering{}%
\begin{tabular}{@{}llr@{~$\pm$~}lr@{~$\pm$~}lr@{~$\pm$~}lccc}
\hline 
Source  & NG~model  & \multicolumn{2}{c}{$A_{1}$} & \multicolumn{2}{c}{$A_{2}$} & \multicolumn{2}{c}{$A_{3}$} & No of  & rms  & interval of data\tabularnewline
 &  & \multicolumn{6}{c}{NG~parameters (Eq.~\ref{eq:ng_std}) in units of $10^{-8}$} & obs.  & {$^{\prime\prime}$}  & {{[}yyyymmdd{]}}\tabularnewline
$[1]$  & $[2]$  & \multicolumn{2}{c}{$[3]$} & \multicolumn{2}{c}{$[4]$} & \multicolumn{2}{c}{$[5]$} & $[6]$  & $[7]$  & $[8]$ \tabularnewline
\hline 
\multicolumn{11}{c}{~~~~}\tabularnewline
\multicolumn{11}{c}{C o m e t ~~~~~C/2007 W1 ~~~B o a t t i n i}\tabularnewline
present  & PRE  & 1.002  & 0.139  & -0.7253  & 0.0032  & -0.4916  & 0.0703  & 926  & 0.49  & 20071120--20080612 \tabularnewline
 & POST  & 5.866  & 0.272  & -0.783  & 0.172  & 0.138  & 0.250  & 777  & 0.59  & 20080630--20081217 \tabularnewline
NK1731A  & PRE  & 1.905  & 0.072  & -0.5243  & 0.0302  & \multicolumn{2}{c}{--} & 804  & 0.57  & 20071120--20080612 \tabularnewline
NK1731B  & POST  & 5.753  & 0.111  & -0.705  & 0.150  & \multicolumn{2}{c}{--} & 733  & 0.63  & 20080630--20081107 \tabularnewline
\multicolumn{11}{c}{~~~~}\tabularnewline
\multicolumn{11}{c}{C o m e t ~~~~~C/2008 A1 ~~~M c N a u g h t}\tabularnewline
present  & NG  & 5.150  & 0.032  & 0.9915  & 0.0201  & 0.1939  & 0.0076  & 997  & 0.44  & 20080110--20100117 \tabularnewline
 & PRE  & 4.608  & 0.136  & 1.894  & 0.233  & 1.844  & 0.203  & 393  & 0.28  & 20080110--20080928 \tabularnewline
 & POST  & 10.094  & 0.282  & 6.142  & 0.291  & -4.431  & 0.306  & 544  & 0.54  & 20081001--20100117 \tabularnewline
NK1807  & NG  & 5.099  & 0.047  & 0.7641  & 0.0272  & \multicolumn{2}{c}{--} & 869  & 0.71  & 20080110--20090714 \tabularnewline
\hline 
\end{tabular}
\end{table*}

Comets C/2007 W1 (1.2 yr of data) and C/2008 A1 (2 yr) exhibit the
most manifesting NG~effects in their motion among comets studied
in this paper. NG~models based on the entire set
of positional data proved to be completely inappropriate for both
these objects; the detailed discussion of NG~models based on full
data sets was given in Section~\ref{sub_orbit_full_data}. Moreover,
NG~effects appear to be variable inside the interval covered by positional
data. Such an erratic behaviour can be detected by using data taken
before perihelion passage and after perihelion to determine the set
of NG~parameters, separately for both orbital branches. The results
are given in Table~\ref{tab:NGTwoComets}, where also the NG~parameters
derived by \citet{NK1731A,NK1731B,NK1807} are shown. One can see
that our values of NG~parameters are in very good agreement with
Nakano though he assumed that $A_{3}=0$ and used slightly different
sets of data. Unfortunately, Nakano analysed NG~effects for pre-perihelion
and post-perihelion data separately only for C/2007~W1, so solely
solutions for this comet could be compared in Table~\ref{tab:NGTwoComets}.

\subsubsection*{C/2006~K3~McNaught}

NG~effects easily determinable from the entire interval of data despite a moderate perihelion distance of 2.5\,au. 
The radial component of NG~acceleration dominates and is well-determined, we decided to include the normal component to the model since this gives slight improvements in rms and O-C-diagram and generalize the NG~solution;
no other tailored model is needed for this object.

\subsubsection*{C/2006~P1~McNaught}

This comet ($q_{\rm osc}=0.17$\,au) was the second brightest comet
observed by ground-based observers since 1935 and demonstrated a spectacularly
structured huge dust tail (e.g. \citealp{Jones:2008a}). It might
be surprising that in case of such an active comet with extremely
small perihelion distance it was possible to obtain a very well determined,
standard NG~orbit from the whole data set. Best NG~solution is based on radial and transverse components of NG~acceleration.

\subsubsection*{C/2007~W1~Boattini }

This comet is also discussed at the beginning of this section together with comet C/2008~A1.
We detected strong and variable NG effects in its
motion. As a result we recommend separate, nongravitational PRE and
POST models for studies of its past and future dynamics. The
similar approach was proposed by Nakano, see Table~~\ref{tab:NGTwoComets}
for more details. Our all osculating orbit solutions 
(Section~\ref{sub_NGmotion}, Tables~\ref{tab:NG-parameters}-\ref{tab:models}) show
that C/2007~W1 ($q_{\rm osc}=0.85$\,au) , having a negative value
of $1/a_{\rm ori}$ is an excellent candidate to be an interstellar
comet. It is therefore important at this moment to refer to two quite
different publications on this comet. In the first, \citet{Villanueva:2011a}
measured a chemical composition of C/2007~W1 using NIRSPEC at Keck-2.
They derived the abundance ratios of eleven volatile species relative
to the water and concluded that almost all these ratios are among
the highest ever detected in comets (see figure~8 therein). Thus,
this comet seems to be very peculiar also in the light of chemical
composition.

\noindent In the second paper interesting from our point of view,
\citet{Wiegert:2011a} reported a new daytime meteor shower detected
using Canadian Meteor Orbit Radar. They analysed the data in the 2002-2009
interval and detected Daytime Craterid shower in two years: in 2003
and 2008. Next, they concluded that this shower can be connected with
C/2007~W1 because of the similarity of both sets of orbital elements,
excluding eccentricities. They argued that the eccentricity of C/2007~W1
is known with so great uncertainty that this comet can be a short-period
comet giving two showers. According to the authors, the second shower
detected in 2008 would be after C/2007~W1 perihelion passage in 2007,
the first one -- at the previous perihelion passage of this comet.
In our opinion, however, the orbit of this comet is known much more
accurately than \citet{Wiegert:2011a} argued. Thus, only the shower
in 2008 could be related to C/2007~W1.

\subsubsection*{C/2007~W3~LINEAR}

NG~effects clearly seen in the motion of this comet
and the standard NG model is easily determinable from the entire
interval of data. The asymmetric model is marginally determinable, however with large uncertainties of 
$\tau$-shift and with no improvements of orbital fitting (Table~\ref{tab:NG-parameters}). No other tailored model is necessary
to represent the positional data of this object.

\subsubsection*{C/2008~A1~McNaught}

This comet is also discussed at the beginning of this section together with comet C/2007~W1.
Due to the nature of the detected NG forces (strong
and variable) we recommend separate, nongravitational PRE and POST
models for studies of its past and future dynamics. For comparison
we show also two models based on the entire data set: an asymmetric
one and a standard, symmetric model.

\subsubsection*{C/2009~K5~McNaught}

This comet also seems to be a peculiar object. Considering rather
long time interval of observations (2.5 yr) it could be optionally
included into the group of comets with long sequences of data since
its observations cover quite large heliocentric distances from 4.35\,au
before perihelion to 6.25\,au after perihelion (orbit 1a+ class).
This fact, together with small perihelion distance, should create
a perfect opportunity to determine the NG~effects either from the
whole data set (as for all comets of long data sets in this paper,
previous subsection) or from pre-perihelion and post-perihelion data
individually, as in the case of C/2007~W1 or C/2008~A1. In contrast
to the expectation, the NG~effects cannot be reliably determined
from the entire data set of C/2009~K5 as well as individually from
pre- or post- perihelion orbit branches. Thus, we decided to include
this comet to peculiar objects. We recommend separate
gravitational PRE and POST models for this comet and present two other
models for comparison.

\subsubsection*{C/2009~R1~McNaught}

This comet of a very small perihelion distance  ($q_{\rm osc}=0.405$\,au)
was observed only prior to its perihelion passage and was lost soon
after it. There is no information about this comet after perihelion
and we can speculate that this comet has disintegrated. Among comets
observed only before perihelion passage C/2009~R1 exhibits strong and well-determinable
NG~effects during interval covered by positional measurements and
belongs to comets with good quality NG~orbit (see also Fig.~\ref{fig:OC_09r1}).

\subsubsection*{C/2010~X1~Elenin}

Another peculiar comet of a very small perihelion distance ($q_{\rm osc}=0.482$\,au).
It starts to disintegrate about one month before perihelion and it
turn out that only a pure gravitational orbit can be well determined
from the shorter interval of data -- the part of data not included
in the orbit determination is shown in light grey in Fig.~\ref{fig:DataDistPeculiar}.
On the occasion of this comet, it is worth mentioning that even when
a cometary disintegration was observed, some authors derived NG~orbit
with standard (constant!) NG~parameters A$_{1}$, A$_{2}$, A$_{3}$,
although they describe the systematic acceleration acting on a comet,
which is a function of the heliocentric distance from the Sun. These
standard NG~parameters cannot correctly account for a sudden change
in orbital motion due to comet's partial disruption. Therefore, the
interpretation of results obtained in such a case should be restricted
to the statement that some NG~effects are clearly seen in the cometary
motion but nothing more. It seems to us that the values of orbital
elements in such a NG~case also should be treated with great caution.
Comet C/2010~X1 just may be an good example of such a case of misusing
the standard MSY~method. 

\subsection{Group C: comets of non-detectable or very weak NG~effects}
\label{sub_cometary_case_C}

\begin{figure}
\includegraphics[width=8.8cm]{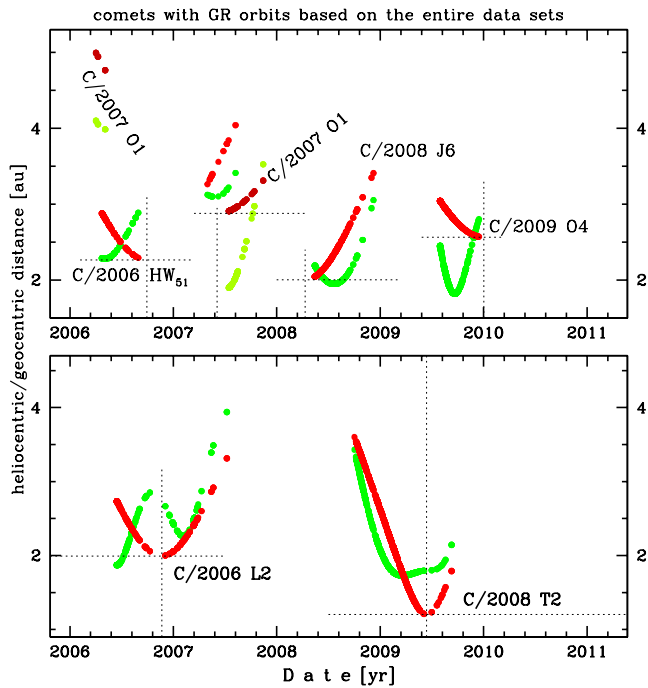} 
\caption{The same as in Fig.~\ref{fig:DataDistLong} for comets of non-detectable
or very weak NG~effects. The horizontal axes covers exactly the same
time-interval in both panels and the same time-scale was used as in
Fig.~\ref{fig:DataDistLong}. }

\label{fig:DataDistGrav} 
\end{figure}

Whether NG~effects are noticeable in comet's motion or not depends
on many factors such as quality and structure of data, the general
level of activity and physical properties of comet (the nucleus structure,
chemical composition, its shape and mass). Thus, generally each case
should be individualized. However, it turns out that we often can
pretty well predict whether it is possible to determine the NG~orbit
from the inspection of structure of the data, where by 'structure'
we mean here all that can be seen in the plot of the heliocentric
and geocentric distances of all positional measurements.

It is rather not surprising that for three of six comets to be described
in this section, namely for C/2007~O1 ($q_{{\rm osc}}=2.88$\,au),
C/2009~O4 ($q_{{\rm osc}}=2.56$\,au) and C/2008~J6 ($q_{{\rm osc}}=2.00$\,au)
we do not succeed in determining NG~effects in their orbital motion.
We can expect this from quick inspection of Fig.~\ref{fig:DataDistGrav}. 

In the remaining three cases, the situation is not as clear as above.
From a review of Fig.~\ref{fig:DataDistGrav} (notice that the scale
of horizontal and vertical axes in both panels are the same) it is
not obvious that NG~effects are not detectable within time intervals
covered by data. Of course, in this type of a qualitative discussion
we assume some physical similarities between considered comets, mainly
in their global activity. 

In fact, for three comets described below, C/2006
HW$_{51}$, C/2006~L2 and C/2008~T2, we detected some traces of
NG~effects in positional data with a negative radial component of
the NG~acceleration (models marked as NG$_{A1}$ in column {[}2{]}
of Table~\ref{tab:models}; see also discussion in Section~\ref{sub_NGmotion}).
However, in all these cases we noticed only slight improvements in
data fitting in comparison to GR~orbit. Therefore, the interpretation
of these NG~models can be twofold. These models reflect the actual
NG~acceleration of these comets (for example giving some indication
of the existence of active sources on the nucleus as was mention above),
or the observed slight improvements in data fitting are only the result
of a larger number of parameters taken into consideration when determining
the NG~orbit. We decided, however, to include these NG~models to
Table~\ref{tab:models} solely as alternative models for dynamical
status discussions based on the previous perihelion calculations,
see Part~II.

\subsubsection*{C/2006~HW$_{51}$~Siding~Spring}

The data structure of this comet is qualitatively quite similar to
that of comet C/2006~K3, a number of measurements is also very similar
(about 200 observations in both cases). Furthermore, both comets passed
perihelion rather far from the Sun (more than 2.2\,au). It seems,
however, that the longer time interval of data (1.7 years), resulting
in a wider range of observed heliocentric distances before perihelion
in the case of C/2006~K3 (3.95\,au at the moment of discovery in
comparison to 2.87\,au at for C/2006~HW$_{51}$), 
causes that the NG~acceleration in C/2006~K3 is clearly visible
in its motion, while in the motion of comet C/2006~HW$_{51}$ it
is not so easily discernible. The NG~effects can be firmly detected
in the motion of C/2006~K3 also due to the fact that this comet was
significantly more active than C/2006~HW$_{51}$. One can speculate
that the activity of the comet C/2006~HW$_{51}$ is limited only
to some active areas somehow specifically located on the surface of
the comet causing that standard MSY~model gives some traces of NG~effects
with negative radial component of NG~acceleration.
We recommend pure gravitational model for this object, but we also
present NG and gravitational PRE models for comparison.

\subsubsection*{C/2006~L2~McNaught}

Similar arguments to the presented above seems to
be correct in the case of comet C/2006~L2, that also passed through
perihelion not very close to the Sun ($q_{\rm osc}=1.994$\,au)
and displays some traces of NG~effects with negative
$A_{1}$. We recommend pure gravitational model for
this object, while NG and gravitational PRE models are presented for comparison.

\subsubsection*{C/2007~O1~LINEAR}

In the data of C/2007~O1, we have more than one-year
gap in positional measurements. This is due to the fact that the object
was initially discovered as an asteroid, and more than a year later
it was rediscovered as a comet. Data of such an unusual structure,
where 12~observations were collected far before perihelion when the
object was more than 4\,au from the Sun and the rest of data were
taken after the perihelion passage, together with the rather large
perihelion distance, can effectively prevent the determination of
NG~effects despite quite a long period formally covered by measurements,
so we present only pure gravitational solution for this comet.

\subsubsection*{C/2008~J6~Hill}

Comet C/2008~J6 was followed from a heliocentric distance of 2.04\,au
to 3.43\,au, thus in a wider range of heliocentric distances what
is more promising. However, it was discovered after passing through
the perihelion. In this case, the chance to detect the NG~acceleration
in the motion of even moderately active comet is low,
only GR model is presented.

\subsubsection*{C/2008~T2~Cardinal}

Comet C/2008~T2 seems to be a more unusual object. In terms of the
structure of data we have a situation quite similar to that of comet
C/2007~W3 (Fig.~\ref{fig:DataDistPeculiar}), but here we have much
more measurements as well as the comet came closer to the Sun at perihelion
(1.20\,au compared to 1.78\,au for C/2007 W3). Both facts should
allow us to determine the NG~effects easier for C/2008~T2 than for
C/2007~W3. However, in this comet the NG~effects are loosely detectable
(almost at a noise level. One can suppose that
comet C/2008~T2 probably exhibits another character of activity than
C/2007~W3 or/and physically differs from C/2007~W3.
We recommend GR model obtained from the entire dataset for this object,
but we also present NG and gravitational PRE models for comparison.

\subsubsection*{C/2009~O4~Hill}

Comet C/2009~O4 was observed only before perihelion passage in the
narrow range of heliocentric distance from 3.04\,au to 2.57\,au
and only during 4.5 months period, therefore only
the gravitational model can be obtained.

\subsection{Group D: comets of weak quality of osculating orbits}
\label{sub_cometary_case_D}

\begin{figure}
\includegraphics[width=8.8cm]{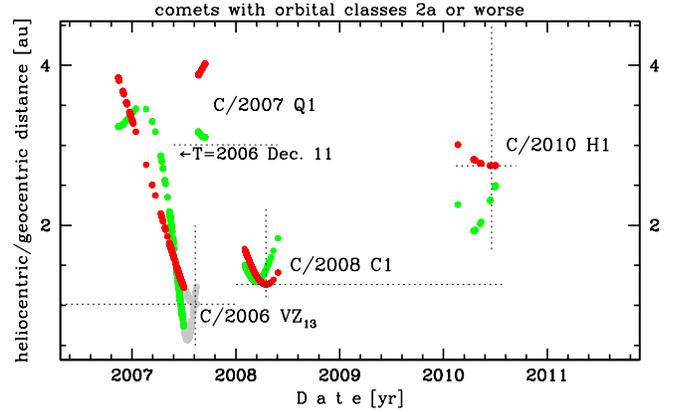} 
\caption{The same as in Fig.~\ref{fig:DataDistLong} for comets of weak quality
of osculating orbits; also the same time-scale was used in horizontal
axis. The time interval not taken for the determination of PRE type
of osculating orbit of comet C/2006~VZ$_{13}$ is shown in light
grey ink.}
\label{fig:DataDistClass2} 
\end{figure}

Due to a generally poor quality of osculating orbits of these four
comets in comparison to others and their asymmetric distribution of
observations relative to perihelion we should be very careful when
making statements about their past and future motion. Moreover, we
should admit that one of them (C/2006 VZ$_{13}$), split into separate
fragments or disintegrated near perihelion.

\subsubsection*{C/2006~VZ$_{13}$~LINEAR}

The observations of C/2006~VZ$_{13}$ ($q_{\rm osc}=1.01$\,au,
the smallest perihelion distance in this group) were stopped very
soon after its perihelion passage. C/2006~VZ$_{13}$ (Fig.~ \ref{fig:DataDistClass2}),
was observed longer than remaining objects in this group, about eight
months, while three others less than five months. However, it belongs
to this group because in fact the only adequate orbit for its past
dynamical evolution can be determined from the pre-perihelion data.
We decided to cut the pre-perihelion string of data on July 1, 2007
($1.23$\,au from the Sun), i.e. 40 days prior to the perihelion
passage because with such a restriction we derived the NG~osculating
orbit that gives O-C~diagram free from any trends in right ascension
or declination. For this reason the orbit based on this time interval
is the most appropriate as starting orbit for the past dynamical evolution
(see Part~II of this investigation). The range of data that was not
used for PRE type of model determination is shown in light grey ink
in Fig.~ \ref{fig:DataDistClass2}. Thus, the data time interval
taken for the past evolutionary studies was only $\sim$7.5 months
in this case. Shortening the time interval of data by almost 20~per
cent resulted here in a more than four-fold reduction of a precision
of $1/{\rm a}_{\rm osc}$-determination, and resulted in a decrease
in its orbital class from 1b (NG orbit determined from the entire
data set) to class 2a (NG~orbit based on pre-perihelion data). For
a future dynamical evolution we have no choice and for this purpose
the NG~orbit based on the entire data set was used. It is worth noting
that the future orbit should be treated with great care because of
our ignorance of the fate of this comet shortly after perihelion passage
(the last observation was taken four days after perihelion).

\subsubsection*{C/2007~Q1~McNaught}

This comet have the greatest perihelion distance ($q_{\rm osc}=3.01$\,au)
and the worst quality orbit (3a) in this group (and in the whole sample
of comets examined in this paper) is determined for
this comet.Its poor quality is a direct consequence of the shortest data arc throughout
the sample (only 24 days) and also of an unusual moment of discovery,
more than eight months after it passed through perihelion. Usually,
when the astrometric observations include perihelion then the orbit
have a chance to be more precisely determined. Only the pure gravitational orbit 
can be determined in this case.

\subsubsection*{C/2008~C1~Chen-Gao}

Despite a very short time interval of data, the NG~effects
are detectable in the motion of this comet ($q_{\rm osc}=1.26$\,au).
However the NG parameters are not well determined and the decrease
in rms is not observed, so we present the NG model for this comet
only for comparison and recommend the GR model.

\subsubsection*{C/2010~H1~Garradd}

It is impossible to detect the NG~effects from the
set of data for this comet due to very narrow data range -- observations
span a short time period. Additionally, we can notice that it passed
the perihelion at the moderately large distances from the Sun ($q_{\rm osc}=2.745$\,au),
and only a small number of measurements were taken. As a result only
a GR model can be obtained.

\section{Original and future orbits}

\label{sec_orbit_original_future}

In the present numerical calculations, a dynamical evolution investigation
of a given object starts from the swarm of VCs constructed using the
osculating orbit (so-called nominal osculating orbit) determined in
the respective model shown in Table~\ref{tab:models}. We performed
dynamical calculations for each model presented in this table. Of
course, for models based on PRE data we follow only the past evolution,
whereas for models based on POST data -- only the future evolution.
Each individual swarm of starting osculating orbits is constructed
according to a Monte Carlo method proposed by \citet{sitarski:1998},
where the entire swarm fulfil the Gaussian statistics of fitting to
positional data used for a given osculating orbit determination. Similarly
to our previous investigations (see for example Paper~1), each swarm
consists of 5\,001\,VCs including the nominal orbit; we checked
that the number of 5\,000 orbital clones gives a sufficient sample
for obtaining reliable statistics at each step of our study, including
the end of our numerical calculations, i.e. at the previous and next
perihelion passage (see Part~II of this investigation). Therefore,
we are able to determine the uncertainties of original and future
reciprocal of semimajor axis ($1/a_{\rm ori}$ and $1/a_{\rm fut}$),
that are here taken at 250\,au from the Sun, i.e. where planetary
perturbations are already completely negligible \citep{todorov-juch:1981}.

Values of $1/a_{\rm ori}$ and $1/a_{\rm fut}$ and their uncertainties
derived by fitting the 1/a-distribution of original and future swarm
of VCs to Gaussian distribution are given in columns $[9]$--$[10]$
of Table~\ref{tab:models}. All $1/a$-distributions as well as distributions
of other orbital elements of analysed comets were still perfectly
Gaussian at 250\,au from the Sun. However, further evolution under
the Galactic tides and stellar perturbation can potentially introduce
significant deformations in the initially 6D-normal distribution of
orbital elements in the swarms of VCs (up to 10D-normal distribution
in the NG~case), what will be shown in the Part~II of this investigation.

\begin{figure}
\includegraphics[width=8.8cm]{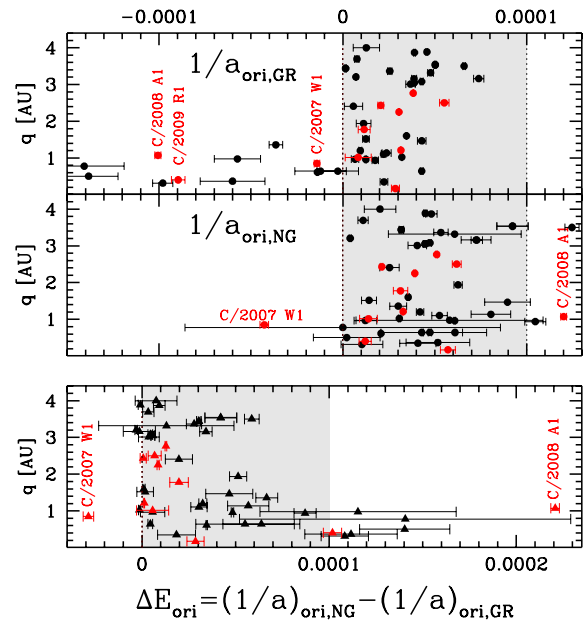}
\caption{Shifts of $1/{\rm a}_{\rm ori}$ due to the
NG~acceleration for eleven comets from the period 2006-2010 (red
symbols) with well-determined NG~effects and 37~previously analysed
comets (black symbols; Paper~1 \& 2). Three largest uncertainties of
$1/{\rm a}_{\rm ori,NG}$ belong to comets C/1959~Y1, C/1952~W1 and
C/1892~Q1. } \label{fig:spik_ng}
\end{figure}

\subsection{Original semimajor axes of NG~orbits}\label{sub_orbit_original_axis_NG}

First, it is important to notice that an osculating NG~orbit
determined from a given set of data is not the same orbit as one
determined from these same observations but under the assumption of
purely gravitational motion \citep{krolikowska:2001}. Therefore, the
change of original $1/a$-value due to incorporating the
NG~acceleration in the process of orbit determination can be even
very large though the NG~effects provide only modest changes in the
original 1/a-value along an individual orbit. Typically, the
original eccentricity of NG~orbit are smaller than eccentricity of
GR~orbit for a given comet with determinable NG~acceleration (where
both GR and NG orbits are derived from the same set of positional
data). We confirmed these general findings in the present studies.
However, we found some interesting atypical behaviour for comet
C/2007~W1.

The differences between the inverse original semimajor axes derived
in NG~model of motion and less realistic pure gravitational model of
motion are presented for 11~comets examined here (red symbols) and
37~comets studied in the Paper1\&2 (black symbols) in
Fig.~\ref{fig:spik_ng}. In the bottom panel the dependence of the
strength of NG~forces on the osculating perihelion distance,
$q_{\rm osc}$ is clearly visible for the whole set of 48~comets
with determined NG~effects.

To maintain consistency with previous our studies, the values of
1/a$_{\rm ori}$ and their uncertainties presented in this plot are
based on osculating NG~orbits determined from the entire data sets
except of the comet C/2007~W1 where we took the value of 1/a$_{\rm
ori}$ determined from the NG~orbit based on pre-perihelion subset of
data as significantly more adequate for this comet (see also
Section~\ref{sub_cometary_case_B}).  \citet{NK1731A,NK1731B} also
considered the pre-perihelion and post-perihelion orbital branches
independently for this particular comet. He derived a similar value
of 1/a$_{\rm ori}=-0.000059$\,au$^{-1}$ on the basis of
pre-perihelion data only (the detailed comparison of both NG~models
is given in Section~\ref{sub_cometary_case_B}. However, for the
NG~orbit based on the entire observational interval we obtained even
a more negative value of 1/a$_{\rm
ori}=-0.000082\pm0.000003$\,au$^{-1}$. Thus, C/2007~W1~Boattini is
the first serious candidate for interstellar comet among LPCs
examined by us so far.

\begin{table*}
\caption{\label{tab:Comets108_threesamples} Three samples of comets
with $1/a_{\rm ori}<10^{-4}$\,au$^{-1}$: qS, qM and qL in
comparison to MWC\,08. The number of comets are given for first
quality class orbits in MWC\,08, except the sample of comets
investigated in this paper, where 3 objects with quality class 2 and
3 are included; we indicate this by '$+$ 3' in columns [3] \& [9]. }
\centering{}%
\begin{tabular}{ccccccccc}
\hline
 & \multicolumn{8}{c}{ N u m b e r ~~~o f ~~~c o m e t s ~~~i n ~~~t h e ~~~d
i f f e r e n t ~~~s a m p l e s}\tabularnewline Period  &
\multicolumn{2}{c}{$q_{{\rm osc}}<3.1$\,au} & \multicolumn{2}{c}{$3.1\le q_{{\rm osc}}<5.2$\,au} & \multicolumn{2}{c}{$q_{{\rm osc}}\ge5.2$\,au} & \multicolumn{2}{c}{All comets}\tabularnewline of discovery  &
MWC\,08  & Sample qS  & MWC\,08  & Sample qM  & MWC\,08  & Sample qL & MWC\,08  & analysed by us\tabularnewline 
{$[1]$}  & {$[2]$}  & {$[3]$}  & {$[4]$}  & {$[5]$}  & {$[6]$}  & {$[7]$}  & {$[8]$}  & {$[9]$} \tabularnewline 
\hline 
before 1900  & 10  & 2  & -- & --  & --  & --  & 10  & 2  \tabularnewline 
1900--1949   & 26  & 3  & 6  & --  & --  & --  & 32  & 3  \tabularnewline
 1950--1999  & 24  &16  & 22 & 17  & 11  & 11  & 57  & 44 \tabularnewline
 2000--2005  & 10  & 6  & 14 & 14  & 12  & 11  & 36  & 31 \tabularnewline
 2006--2010  & 12  & 19 $+$ 3  & 3  & 4  & 3  & 2  & 18  & 25 \tabularnewline 
\hline 
 All         & 82  & 46 $+$ 3  & 45  & 35  & 26  & 24 & 153  & 105 $+$ 3 \tabularnewline 
\hline
\end{tabular}
\end{table*}

\noindent In the case of comet C/2008~A1 we show in Fig.~\ref{fig:spik_ng}
the results based on the entire data interval though for this comet
the NG~orbit determined from the full interval is also highly unsatisfactory
(as we discuss earlier). The largest differences in $\Delta{\rm E}_{\rm ori}$
was derived just for this comet (the rightmost red symbol in the bottom
panel of Fig.\ref{fig:spik_ng}), whereas based on a pre-perihelion
subset of data we derived 1/a$_{\rm ori}=0.000120\pm0.000002$\,au$^{-1}$
for NG~orbit and $\Delta{\rm E}_{\rm ori}=-0.000011$\,au$^{-1}$.
However, we would like to stress that C/2008~A1 as well as C/2007~W1
are comets that both have extremely strongly manifesting and variable
NG~effects inside their observational intervals.

Sets of original and future 1/a-values taken for statistical
analysis (see next section) are based on the preferred osculating
orbits. It is obvious that NG~orbits -- if determinable in the
motion of a given comet -- are always more realistic than pure
gravitational orbit derived from the same data set.

\subsection{Observed distributions of original and future semimajor axes}
\label{sub_orbit_original_future_statistics}

In the discussion below we do not pretend to describe
planetary perturbations on LPCs in general. Instead, we aim to present
really observed changes in $1/a$ for a significant percentage of
the observed LPCs, additionally for the first time fully accounting
for their uncertainties. We also discuss the contribution of the NG~forces 
into the $1/a$ change during a cometary flyby through the
planetary system.

To describe the original and future $1/a$-distributions
we divided 108~comets investigated by us so far 
(in the present investigation and Papers~1 \& 2) into three subsamples 
of a different completeness according to their osculating perihelion distances: 
\begin{description}
\item [qS] -- this sample consists of 49\,LPCs with $q_{\rm osc}<3.1$\,au
and includes comets with strongly manifesting NG~effects in their
motion discovered before 2006 (Papers 1 \& 2) and a
complete sample of comets discovered in the period 2006-2010 investigated
in this work. Thus, we have here 46 of 89 (52 per cent) comets
discovered before 2011 with first quality class orbit by using original
MSE method (Section~\ref{sec_orbit_accuracy}) and additionally three
comets of poorer quality class orbits. 
\item [qM] -- complete sample of 35\,LPCs with 3.1\,au\,$\le q_{\rm osc}<5.2$\,au
discovered in the period 1970-2006 (first quality class orbits, Paper~2).
\item [qL] -- almost complete sample of 24\,LPCs with $q_{\rm osc}\ge5.2$\,au
discovered before 2008 (first quality class orbits, Paper 2); only
two objects, C/2005~L3 and C/2007~D1, were not taken into account
in Paper~2 because they were still observable at the beginning of
2012. 
\end{description}

\begin{table*}
\caption{\label{tab:Comets_NG_models_future_original} Contribution of planetary
perturbations and NG~effects to $\delta(1/a)=1/a_{\rm fut}-1/a_{\rm ori}$
for nine comets with NG~orbits given as the preferred models of their
motion in Table~\ref{tab:models}. Comets are arranged in order of
increasing $q_{\rm osc}$. In the first row of individual comet
$\delta(1/a)$ for the main NG~model (given in the first rows in
Table~\ref{tab:models}) is presented, the second row shows $\delta(1/a)$
obtained using the best GR~model and third row displays the $\delta(1/a)$
derived using the same NG~orbit as in the first row, however during
integration backward and forward to the 250\,au from the Sun the
NG~acceleration was excluded. In the second row (column $[9]$) the
percentage change in $\delta(1/a)$ relative to first row is also
given. Values of $1/a_{\rm ori}$, $1/a_{\rm fut}$, and $\delta(1/a)$
are given in units of $10^{-6}$\,au$^{-1}$ used in this paper,
and the magnitude of the NG~acceleration, $A=\sqrt{A_{1}^{2}+A_{2}^{2}+A_{3}^{2}}$,
in units of $10^{-8}$\,au$\cdot$day$^{-2}$.}

\centering{}%
\begin{tabular}{ccccrrcrlrrrr}
\hline 
Comet  & $q_{{\rm osc}}$  & type of  & Model of  & rms  & $1/a_{{\rm ori}}$  & Model of  & $1/a_{{\rm fut}}$  & $\delta(1/a)$  & \multicolumn{4}{c}{N G~~~p a r a m e t e r s}\tabularnewline
 & {[}au{]}  & motion  & orbit for  & for $[4]$  &  & orbit for  &  &  & \multicolumn{2}{c}{pre-perihelion} & \multicolumn{2}{c}{post-perihelion}\tabularnewline
 &  &  & $1/a_{{\rm ori}}$  & $^{\prime\prime}$  &  & $1/a_{{\rm fut}}$  &  &  & $A$  & $A_{2}/A_{1}$  & $A$  & $A_{2}/A_{1}$ \tabularnewline
{$[1]$}  & {$[2]$}  & {$[3]$}  & {$[4]$}  & {$[5]$}  & {$[6]$}  & {$[7]$}  & {$[8]$}  & {$[9]$}  & {$[10]$}  & {$[11]$}  & {$[12]$}  & {$[13]$} \tabularnewline
\hline 
C/2006 P1  & 0.171  & NG  & NG  & 0.25  & 57.2  & NG  & 467.6  & ~410.5  & 1.33  & 0.23  & 1.33  & 0.23 \tabularnewline
 &  & GR  & GR  & 0.25  & 31.4  & GR  & 490.7  & ~459.3 (11.4\%)  &  &  &  & \tabularnewline
 &  & GR  & NG  & 0.79  & 37.4  & NG  & 496.7  & ~459.3  &  &  &  & \tabularnewline
C/2009 R1  & 0.405  & NG  & NG  & 0.51  & 12.2  & NG  & 170.4  & ~158.3  & 6.02  & 0.24  & 6.02  & 0.24 \tabularnewline
 &  & GR  & GR  & 0.63  & -89.6  & GR  & -723.0  & -812.6  &  &  &  & \tabularnewline
 &  & GR  & NG  & 4.87  & 12.2  & NG  & -592.4  & -551.7  &  &  &  & \tabularnewline
C/2007 W1  & 0.850  & NG  & PRE,NG  & 0.49  & -42.7  & POST,NG  & 554.4  & ~586.5  & 1.33  & 0.72  & 5.92  & 0.13 \tabularnewline
 &  & GR  & PRE,GR  & 0.61  & -14.1  & POST,GR  & 892.0  & ~632.9 (7.9\%)  &  &  &  & \tabularnewline
 &  & GR  & PRE,NG  & 4.50  & 71.8  & POST,NG  & 716.5  & ~634.0  &  &  &  & \tabularnewline
C/2006 VZ$_{13}$  & 1.015  & NG  & PRE,NG  & 0.39  & 14.0  & NG  & 491.2  & ~269.3  & 2.13  & 0.46  & 5.63  & 0.66 \tabularnewline
 &  & GR  & GR  & 0.40  & 13.6  & GR  & 442.2  & ~435.6 (62\%)  &  &  &  & \tabularnewline
 &  & GR  & PRE,NG  & 5.31  & 132.8  & NG  & 371.2  & ~~30.4  &  &  &  & \tabularnewline
C/2008 A1  & 1.073  & NG  & PRE,NG  & 0.28  & 120.8  & POST,NG  & 246.5  & -534.1  & 4.98  & 0.41  & 12.62  & 0.61 \tabularnewline
 &  & GR  & PRE,GR  & 0.47  & 132.4  & POST,GR  & 258.4  & ~268.9  &  &  &  & \tabularnewline
 &  & GR  & PRE,NG  & 14.11  & 137.6  & POST,NG  & 1065.05  & ~267.6  &  &  &  & \tabularnewline
C/2007 W3  & 1.776  & NG  & NG  & 0.52  & 31.4  & NG  & 343.9  & ~312.5  & 5.56  & 0.45  & 5.56  & 0.45 \tabularnewline
 &  & GR  & GR  & 0.54  & 14.2  & GR  & 408.8  & ~394.6 (26.3\%)  &  &  &  & \tabularnewline
 &  & GR  & NG  & 3.70  & 21.2  & NG  & 415.8  & ~394.6  &  &  &  & \tabularnewline
C/2006 OF$_{2}$  & 2.431  & NG  & NG  & 0.36  & 21.2  & NG  & -658.8  & -680.0  & 2.75  & 0.58  & 2.75  & 0.58 \tabularnewline
 &  & GR  & GR  & 0.38  & 23.7  & GR  & -662.4  & -686.1 (0.9\%)  &  &  &  & \tabularnewline
 &  & GR  & NG  & 0.49  & 24.9  & NG  & -661.1  & -686.1  &  &  &  & \tabularnewline
C/2006 K3  & 2.501  & NG  & NG  & 0.54  & 61.0  & NG  & -131.3  & -192.3  & 15.85  & 0.14  & 15.85  & 0.14 \tabularnewline
 &  & GR  & GR  & 0.69  & 61.9  & GR  & -123.0  & -184.9 (3.9\%)  &  &  &  & \tabularnewline
 &  & GR  & NG  & 1.19  & 62.0  & NG  & -122.9  & -184.9  &  &  &  & \tabularnewline
C/2006 Q1  & 2.764  & NG  & NG  & 0.37  & 51.1  & NG  & 707.4  & ~656.4  & 35.19  & 0.06  & 35.19  & 0.06 \tabularnewline
 &  & GR  & GR  & 0.50  & 38.3  & GR  & 696.6  & ~658.3 (0.3\%)  &  &  &  & \tabularnewline
 &  & GR  & NG  & 2.05  & 52.7  & NG  & 711.0  & ~658.3  &  &  &  & \tabularnewline
\hline 
\end{tabular}
\end{table*}

\begin{figure}
\includegraphics[width=8cm]{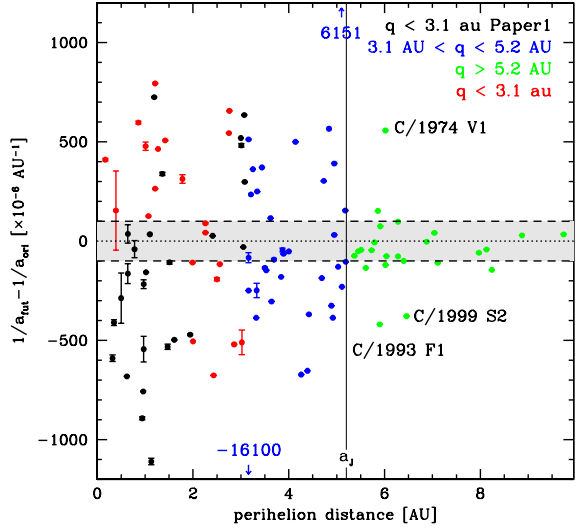} 
\caption{Change of reciprocals of semimajor axis during passage through the
planetary zone, $\delta(1/a)=1/a_{\rm fut}-1/a_{\rm ori}$, as
a function of osculating perihelion distance, where the sample of
22 comets investigated in this paper are shown by red dots. The sample
of 86 comets studied in Paper1\&2 were divided into three subsamples
qS, qM and qL (see Table~\ref{tab:Comets108_threesamples}) and are
given as black, blue and green dots, respectively. The vertical line
marks the position of Jupiter semimajor axis at 5.2\,au. The horizontal grey band between two horizontal dashed
lines shows the area where planetary perturbations are smaller than
$10^{-4}$\,au$^{-1}$.}
\label{fig:Dist_108comets} 
\end{figure}

More detailed descriptions of these three samples are given in Table~\ref{tab:Comets108_threesamples}
where the completeness of these samples in comparison to MWC\,08
is shown. MWC\,08 is complete to the end of the
2007. In the period of 2006--2010, considerable number of 48~LPCs
with $1/a_{\rm ori}<10^{-4}$\,au$^{-1}$ were discovered (eight
of them are still observable), and all of them,
having $q_{\rm osc}<3.1$\,au are investigated in this paper
and are shown by red points in Fig.~\ref{fig:Dist_108comets}. In
total, our sample of 108 comets discussed here constitutes more than
60 per cent of all first class Oort spike comets discovered after
1800 (78 per cent of those discovered after 1950), for which observations
are finished. For the present discussion only the preferred models (shown in the first row for each comet in Table~\ref{tab:models}) were taken into consideration.

In a pure GR~model of cometary motion, the values of $\delta(1/a)=1/a_{\rm fut}-1/a_{\rm ori}$
directly inform about planetary perturbation which a comet suffered
passing through the planetary system. When the NG~osculating orbit
is derived we should expect that NG~effects contribute in some extent
to the value of $\delta(1/a)$. It is not so easy to measure these
contributions, in particular for these comets where two separate NG~osculating
orbits are preferred to describe their actual orbital motion. We have
made such an attempt in the case of comets with NG~orbits investigated
here.

Table~\ref{tab:Comets_NG_models_future_original} presents three
values of $\delta(1/a)$ for all comets with NG~models chosen by us as 
the preferred solution in Table~\ref{tab:models}.
For each comet, the first row gives $\delta(1/a)$-value for the recommended
NG~model, the second row -- $\delta(1/a)$-value for the best GR~model,
whereas the third row shows $\delta(1/a)$-value derived for the same
NG~osculating orbit taken as a starting orbit, however NG~accelerations
were not included during the integration backward and forward to 250\,au
from the Sun for original and future 1/a-determination. Of course,
the last orbit does not represent data (see column $[5]$) since the best 
GR~orbit that gives good data fitting (second row,
column $[4]$) is substantially different than the best NG~orbit.
However, the difference between first row and second (and third) 
row may give some estimation for the NG~contribution
to the $\delta(1/a)$-value. For comets with separate
NG~orbits derived for pre-perihelion branch and post-perihelion branch, 
we calculate the $\delta(1/a)$ from the equation:

\[
\delta(1/a)=(1/a_{\rm fut}-1/a_{\rm osc,post})-(1/a_{\rm ori}-1/a_{\rm osc,pre}),
\]

\noindent where $1/a_{\rm osc,post}$ and $1/a_{\rm osc,pre}$
are values of $1/a_{\rm osc}$ for post-perihelion and pre-perihelion
orbit, respectively.

One can notice using Table~\ref{tab:Comets_NG_models_future_original}
that for comets with perihelion distance greater than 2.0\,au the
$\delta(1/a)$-value can be interpreted as planetary perturbations
even in the NG~models of motion. We have also confirmed
this conclusion for all comets studied in Paper~1 \& 2 where differences
in $\delta(1/a)$-value between NG~and GR~models do not exceed 10
per cent, except C/1997~J2 where we obtained this difference 
at the level of 40 per cent due to a very small value of $\delta(1/a)$
($-41$ and $-30$ for GR and NG~orbit, respectively). However, for
smaller perihelion distances ($q_{\rm osc}<2.0$\,au) we can obtain even 
a change of sign of $\delta(1/a)$, like for C/2009~R1
($q_{\rm osc}=0.405$\,au) and C/2008~A1 ($q_{\rm osc}=1.07$\,au)
with the highest magnitude of NG~acceleration (columns $[10]$--$[13]$
in Table~\ref{tab:Comets_NG_models_future_original}). Differences
of more than 30 per cent in $\delta(1/a)$ between NG and GR models
are found only for three comets investigated here
(two peculiar comets, C/2008~A1 and C/2009~R1, and one of weak quality
of osculating orbit, C/2006~VZ$_{13}$) and five comets of 37 with
NG~orbits determined in Papers~1 \& 2. In the following discussion,
when we use the term 'planetary perturbation', we keep in mind that
identification of planetary perturbation with $\delta(1/a)=1/a_{\rm fut}-1/a_{\rm ori}$
in the case of NG~orbit is subject to an additional contribution,
that exceed 30~per cent of $\delta(1/a)$-value just for a few of
them. This contribution is obviously caused by slightly different
evolution of orbital elements due to NG~acceleration along a given
orbit.

In Fig.~\ref{fig:Dist_108comets} we show $\delta(1/a)=1/a_{\rm fut}-1/a_{\rm ori}$,
in a function of osculating perihelion distance for the sample of
22 comets investigated here (red points) in comparison with all remaining
86~comets investigated by us so far. The uncertainties of $\delta(1/a)$
were derived by calculating the $\delta(1/a_{i})$ where $i=1,...,5001$
for each VC in the swarms, and then fitting the derived $\delta(1/a)$-distribution
of individual swarm of VCs to Gaussian distribution. The uncertainties
of perihelion distance and inclination are significantly below the
size of points in these figures. It is interesting to notice that
the largest $\delta(1/a)$-uncertainty in the group of comets investigated
in the present paper and in the entire sample of 108~Oort spike comets,
were obtained for C/2009~R1 McNaught (orbital class: 1b), easy recognizable
red point with largest vertical error bar. For C/2009~R1 we have
$\delta(1/a)=154\pm199$ in units used in this paper. At a first glance,
this comet with large inclination to ecliptic ($\sim77$\degr) seems
to suffer a relatively moderate planetary perturbations (NG~model).
However, we found that the nominal VC have passed about 1.3\,au from
Jupiter in 2009 (on August 25), and 0.15\,au from Mercury in July
2010 (six days after perihelion passage, the last observation of this
comet was taken 10 days earlier). 

\noindent Among the sample of comets from the 2006--2010 period, the
second comet with large error of $\delta(1/a)$ is C/2007~Q1~Garradd
($\delta(1/a)=-510\pm62$), however, this resulted from the extremely
poor quality of its orbit (class 3a).

\begin{figure}
\includegraphics[width=8.8cm]{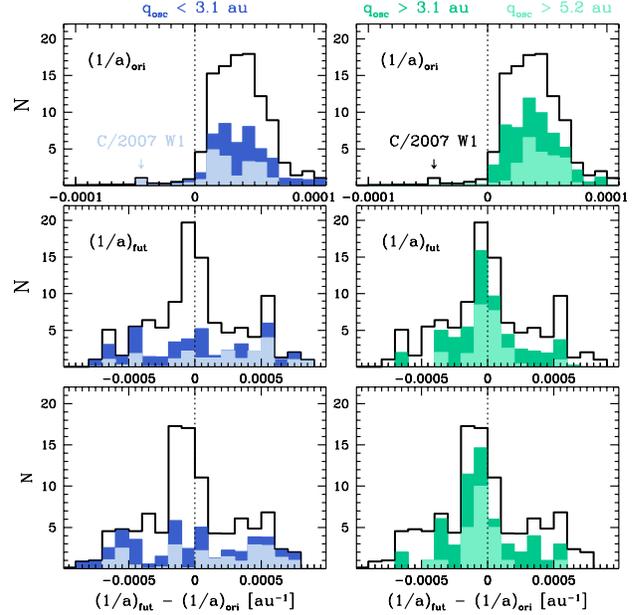} 
\caption{Distribution of $1/a_{\rm ori}$ (top panels), $1/a_{\rm fut}$
(middle panels) and $\delta(1/a)$ for small perihelion ($q_{\rm osc}<3.1$\,au,
left-side filled histograms) and large perihelion ($q_{\rm osc}\ge3.1$\,au,
right-side filled histograms) Oort spike comets. The histogram given
in each panel in thick black ink always represents the distribution
of the whole sample of 108~comets. The uncertainties of $1/a$-determinations
were incorporated into these $1/a$-histograms by taking the full
cloud of VCs for each comet. The light steel-blue histograms in the
left-side panels show the sample of 22~comets investigated in this
paper, where the light turquoise histograms in the right-side panels
represent the sample of comets with $q_{\rm osc}\ge5.2$\,au. Completeness
of all samples are shown in Table~\ref{tab:Comets108_threesamples}.}
\label{fig:OortSpike_smallq_largeq} 
\end{figure}

Apart from three comets (C/1940~R2, C/1980~E1 and
C/2002~A3) lying outside Fig.~\ref{fig:Dist_108comets} all other
suffer only moderate perturbations in $1/a$, with a significant per cent
of very small values (points inside the horizontal band around zero in the figure), conversely to widely spread opinion, that comets coming closer to the Sun than
Jupiter almost certainly suffer strong planetary perturbations. It means, that
in the sample of analysed Oort spike comets the number of objects
that have a chance for returning in the next perihelion passage as
an Oort spike comet is remarkable (see middle panels in Fig.~\ref{fig:OortSpike_smallq_largeq})
and accounts to about 14 per cent of all investigated by us near-parabolic
comets (108~objects). This shows the observed transparency of the
Solar system (see \citealp{dyb-trans:2004}). This transparency seems
to be significantly different for large perihelion near-parabolic
comets than for small perihelion comets and, taking
into account small number statistics fluctuations it is consistent
with the results of the Monte Carlo simulations of this phenomenon,
recently published by \citet{fouchard-r-f-v:2013}.
The detailed percentages of comets with $1/a_{\rm fut}$ still smaller than 
$100\times 10^{-6}$\,au$^{-1}$ within three discussed samples are as follows:

\vspace{0.3cm}

\begin{tabular}{lr}
qS sample~~~~~~  & 10 per cent \tabularnewline
qM sample        &  6 per cent \tabularnewline
qL sample        & 33 per cent. \tabularnewline
\end{tabular}

\vspace{0.3cm}

This feature of the $1/a_{\rm fut}$ distribution is also clearly
visible in Fig.~\ref{fig:Dist_108comets}, where the distributions
of $\delta(1/a)$ as well as the distributions of $1/a_{\rm ori}$
and $1/a_{\rm fut}$ are presented in Fig.~\ref{fig:OortSpike_smallq_largeq}
in the bottom, top and middle panels, respectively. The filled light
steel-blue histograms show the distributions of 22~comets investigated
in this paper. These steel-blue histograms, are overprinted
on the blue histograms that represent qS~sample (49 comets). Similarly,
the distributions of comets of $q_{\rm osc}\ge5.2$\,au (filled
light turquoise histograms, 24 objects) are overprinted on the distributions
of qM$+$qL~sample of comets (filled green histograms, 59 objects
of $q_{\rm osc}\ge3.1$\,au). It should be stressed here that the
uncertainties of $1/a_{\rm ori}$, $1/a_{\rm fut}$ and $\delta(1/a)$
are taken into account in these histograms by considering full cloud
of 5001~virtual orbits for each comet. For example, the global distribution
of $1/a_{\rm ori}$ in both upper panels (marked by thick black
ink) is based on all clouds of VCs of 108~comets, thus in total of
540\,108\,VCs, whereas the global distributions of $1/a_{\rm fut}$
and $\delta(1/a)$ are based on 107 cometary swarms (in total of 535\,107\,VCs)
where C/2010 X1 was not taken into account for future distribution
because of its disruption during perihelion passage. One can see that
a prominent maximum of the observed planetary perturbation
is visible only for comets with $q_{\rm osc}\ge3.1$\,au whereas
for qS sample of comets the distribution of planetary perturbation
is broad with no sharp peak.

There is another interesting feature visible in Fig.~\ref{fig:OortSpike_smallq_largeq}.
The secondary peak between $0.0005$\,au$^{-1}$ and $0.0006$\,au$^{-1}$
is present in the distributions of $1/a_{\rm fut}$ that is visible
in the qS~sample as well as in the sample of comets with $q_{\rm osc}\ge3.1$.
This corresponds to the local maximum between 1\,700\,au and 2\,000\,au
for the future semimajor axes. Outside the value of $0.0006$\,au$^{-1}$
we observe significant and fast decrease in the $1/a_{\rm fut}$-distribution.
It is interesting that the similar trend of decrease is observed around
$0.0006$\,au$^{-1}$ in the $1/a_{\rm ori}$-distribution for
the sample of near-parabolic comets in MWC\,08 (of first quality
orbits).

We noticed that the investigated here sample of small perihelion Oort
spike comets differs from the sample of small perihelion comets with
strongly manifesting NG~effects investigated in Paper~1 in the context
of future history of these objects. Then, we derived that 60 per cent
of them will be lost on hyperbolic orbits in the future, here we have
only 33 per cent of such comets (seven objects escaping on hyperbolic
trajectories and one object that disintegrated at perihelion). Taking
into account all Oort spike comets with small perihelion distances
investigated by us so far (49 objects) we have now 49 per cent of
comets escaping in the future from Solar system on hyperbolas whereas
for the whole sample of comets -- the similar value of 53 per cent.
Thus, the small perihelion cometary sample and the large perihelion
sample appear to be quite similar in this regard. However, more comets,
especially these with $q_{{\rm osc}}<3.1$\,au need to be analysed
with the NG~effects taken into account. In the Part~II we will discuss
the differences between both samples in the context of dynamical status
of near-parabolic comets.

\section{Summary}
\label{sec:Summary}

We showed that the individualized approach to the osculating orbit
determination is advisable, especially for the purpose of past and
future dynamical evolution of near-parabolic comets. Developing such
an attempt we have determined the osculating orbits for a complete
sample of so-called Oort spike comets with a small perihelion distance
($q_{\rm osc}<3.1$\,au) discovered in five year period from 2006
to 2010. For eleven of these comets (50 per cent) we detected NG~effects
in their orbital behaviour using entire sets of data (Table~\ref{tab:Obs-mat}) 
and discussed various NG models (Table~\ref{tab:NG-parameters}). 
The detailed investigation shows, however, that in five of them even
more individualized approach is necessary (Table~\ref{tab:models}).
Thus, we determined osculating orbits for pre-perihelion and post-perihelion
orbital branch, separately. We argued that the separate NG~solutions
for each orbital leg for two of them, C/2007~W1 and C/2008~A1, are
more adequate for past and future orbital evolution. Our solutions
for both comets are very similar to solutions previously obtained
by Nakano (Section~\ref{sub_cometary_case_B}). Additionally, for
C/2006~VZ$_{13}$ the NG~solution based on pre-perihelion orbital
leg is presented. For the remaining two comets with NG~effects detectable
in the entire (very long) data sets, C/2007~N3 and C/2007~Q3, we
proposed GR~orbits derived from a pre-perihelion and post-perihelion
branch as the best osculating orbit for past and future evolutionary
calculations, respectively. Unfortunately, the NG~effects were indeterminable
for the pre-perihelion and post-perihelion leg separately in both
comets despite long-time series of data.

Next, we discuss the possibility of NG~acceleration detection in
the motion of near-parabolic comets in the light of data location
along their orbital tracks.

We also proposed a modified orbital accuracy assessment on the basis
of classical MSE method. Using 22~Oort spike comets investigated
in this paper and 86~comets examined in Papers~1 \& 2 we showed
that the proposed modifications provide a better diversification between
orbital quality classes of currently discovered comets.

Next, we have analysed the original and future inverse semimajor axes
of these 22~near-parabolic comets taken at the distance of 250\,au
before their entrance to the inner Solar system and at the distance
of the 250\,au after perihelion passage, respectively. We discussed
these results in the context of the whole sample of 108, so called,
Oort spike comets investigated by us so far. Our conclusions are as
follows: 
\begin{itemize}
\item We noticed a different shape and slightly different median value of
$1/a_{\rm ori}$-distributions for small perihelion ($q_{\rm osc}<3.1$\,au)
and large perihelion ($q_{\rm osc}\ge3.1$\,au) samples of near-parabolic
comets. In the upper panels in Fig.~\ref{fig:OortSpike_smallq_largeq},
the $1/a_{\rm ori}$-distribution for small perihelion comets is
more broad than for large perihelion comets. However, in these distributions
included are both dynamically new comets as well as dynamically old.
Thus, a more detailed discussion is necessary and this will be given
in Part~II where the results for previous perihelion distance are
presented. Having the value of $q_{\rm prev}$ we can construct
$1/a_{\rm ori}$-distribution for dynamically new and dynamically
old comets separately, and then discuss the shape of actual Oort spike,
i.e. comets first time coming from the Oort spike in the light of
dynamical investigation covering three consecutive perihelion passages
(thus in a scale of tens of millions of years). 
\item Among investigated near-parabolic comets only C/2007~W1 seems to
have an interstellar origin. This, together with \citet{Villanueva:2011a}
findings that the abundance ratios of almost all important volatiles
in C/2007~W1 are one of the highest ever detected in comets, makes
this comet particularly unique. 
\item Prominent maximum of $\delta(1/a)$ is visible only for comets with
$q_{\rm osc}\ge3.1$\,au whereas for small perihelion comets the
distribution of $\delta(1/a)$ is broad with no sharp peak (Fig.~\ref{fig:OortSpike_smallq_largeq}). 
\item The number of comets leaving the Solar system as so called Oort spike
comets is 14\,per cent in the investigated sample (middle panels
in Fig.~\ref{fig:OortSpike_smallq_largeq}) Moreover, about half
of them (7\,per cent of all, 33 per cent of qL sample) are comets
with large perihelion distance ($q_{{\rm osc}}\ge5.2$\,au). To say,
that these comets will be Oort spike comets in the future is rather
premature and the analysis of their future motion and next perihelion
distance is necessary -- this will be given in Part~II of this study. 
\item In the sample of near-parabolic comets with small perihelion distances
investigated by us so far (49 objects, among them 22 investigated
in this paper) we have now 49 per cent of comets escaping in the future
from the Solar system on barycentric, hyperbolic orbits. Since the
similar value of 53 per cent is derived for the entire sample, we
conclude that the small perihelion and large perihelion near-parabolic
samples appear to be similar in this regard. However, more comets,
especially these with $q_{\rm osc}<3.1$\,au need to be analysed
by taking into account NG~effects. 
\item We noticed a secondary peak between $0.0005$\,au$^{-1}$ and $0.0006$\,au$^{-1}$
in the distribution of $1/a_{\rm fut}$ (middle panels in Fig.~\ref{fig:OortSpike_smallq_largeq}).
This corresponds to a local maximum between $\sim$1\,700\,au and
2\,000\,au for the future semimajor axes. Above the value of $0.0006$\,au$^{-1}$
we observe a significant and fast decrease in the $1/a_{\rm fut}$-distribution.
A similar trend of decrease is observed around $0.0006$\,au$^{-1}$
in $1/a_{\rm ori}$-distribution for the sample of near-parabolic
comets in MWC\,08 (of first quality orbits). 
\end{itemize}

\section*{Acknowledgements}

We thank the anonymous referee for the review that improved this paper.
The main part of the orbital calculation was performed using the
numerical orbital package developed by Professor Grzegorz Sitarski
and the Solar System Dynamics and Planetology Group at SRC PAS.

\bibliographystyle{mn2e}
\bibliography{moja22,mkr_22comets}

\label{lastpage}
\end{document}